\documentclass[twocolumn,prl,amsmath,amssymb,longbibliography,floatfix,superscriptaddress]{revtex4-2}
\usepackage{amsmath}
\usepackage{amssymb}
\usepackage{xcolor}
\usepackage{physics}
\usepackage{graphicx}
\usepackage{tabularx}
\usepackage{dcolumn} % Align table columns on decimal point
\usepackage{bm}      % bold math
\usepackage{array}   % table making
\usepackage{amsfonts}
\usepackage{url}
\usepackage[bookmarks=false,colorlinks,citecolor=red]{hyperref}
\usepackage{mathtools}
\usepackage[ruled]{algorithm2e}
\usepackage[capitalize]{cleveref}

\def\>{\rangle}
\def\<{\langle}
\def\Tr{\mathrm{Tr}}

\newcommand{\kket}[1]{|#1\rangle\!\rangle}
\newcommand{\bbra}[1]{\langle\!\langle #1|}

\usepackage{pdfpages} % include pdfs
\usepackage{pgffor} % for loops
\makeatletter
\AtBeginDocument{\let\LS@rot\@undefined}
\makeatother
\def\supplementfilename{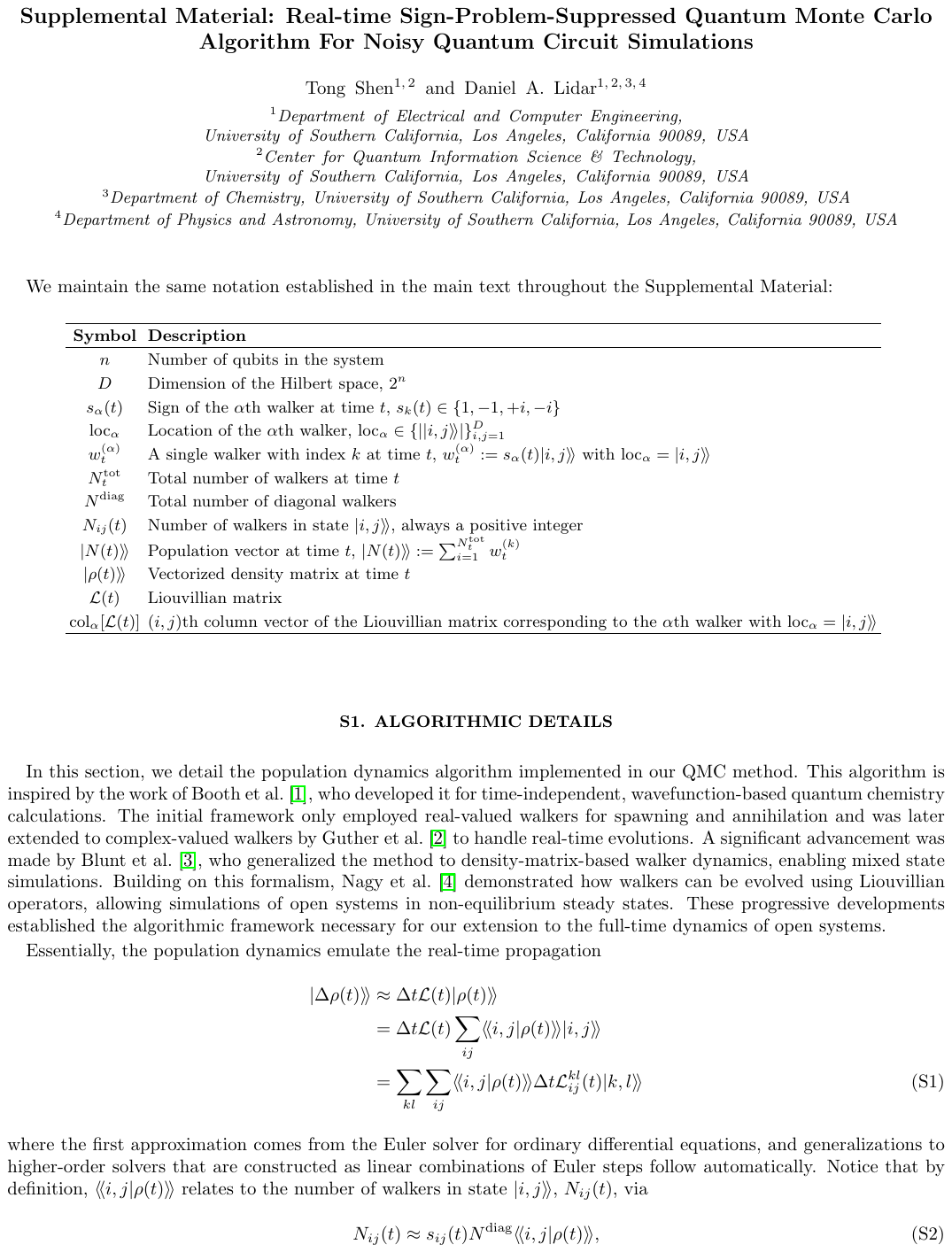}
\pdfximage{\supplementfilename}
\def\numbersupplementpages{\the\pdflastximagepages}

\begin{document}

\include{MyCommand}
\newcommand{\TS}[1]{{\color{blue}{#1}}}
\newcommand{\edit}[1]{{\color{purple}{#1}}}
\newcommand{\DL}[1]{{\color{red}{#1}}}

\title{Real-time Sign-Problem-Suppressed Quantum Monte Carlo Algorithm For Noisy Quantum Circuit Simulations}

\author{Tong Shen}
 \affiliation{Department of Electrical and Computer Engineering, University of Southern California, Los Angeles, California 90089, USA}
 \affiliation{Center for Quantum Information Science \& Technology, University of Southern California, Los Angeles, California 90089, USA}

\author{Daniel A. Lidar}
 \affiliation{Department of Electrical and Computer Engineering, University of Southern California, Los Angeles, California 90089, USA}
 \affiliation{Center for Quantum Information Science \& Technology, University of Southern California, Los Angeles, California 90089, USA}
 \affiliation{Department of Chemistry, University of Southern California, Los Angeles, California 90089, USA}
 \affiliation{Department of Physics and Astronomy, University of Southern California, Los Angeles, California 90089, USA}
 
\begin{abstract}
We present a real-time quantum Monte Carlo algorithm that simulates the dynamics of open quantum systems by stochastically compressing and evolving the density matrix under both Markovian and non-Markovian master equations. Our algorithm uses population dynamics to continuously suppress the sign problem, preventing its accumulation throughout the evolution. We apply it to a variety of quantum circuits and demonstrate significant speedups and scaling improvements over state-of-art quantum trajectory methods and convergence to exact solutions even in non-Markovian regimes where trajectory methods fail. Our approach improves the efficiency of classical simulation of gate-based quantum computing, quantum annealing, and general open system dynamics.
\end{abstract}

\maketitle

As quantum processors continue to improve and grow in scale, the need for numerically efficient classical simulations of the dynamics of quantum circuits becomes more pressing. While ideal circuits follow closed-system models, real-world quantum hardware is inevitably affected by environmental noise, necessitating open quantum system theory~\cite{breuer2002theory, alicki_quantum_2007, weiss2012quantum, rivas_open_2012}. The time-local master equation for a system with density matrix $\rho(t)$ is:
\begin{equation}
\frac{d\rho(t)}{dt} = -i[H(t), \rho(t)] + \frac{1}{2}\sum_k \gamma_k(t) \mathcal{D}[L_k] \rho(t),
\label{eqn:time_local_ME}
\end{equation}
where $H(t)$ represents the system Hamiltonian and driving fields (gates), while $\mathcal{D}[L_k]$, with Lindblad operators $L_k$, describes environmental interactions with $\mathcal{D}[L]\rho(t) = 2L\rho(t)L - \{L^{\dagger}L, \rho(t) \}$ and rates $\gamma_k(t)$. When all rates are positive \cref{eqn:time_local_ME} yields the celebrated Gorini–Kossakowski–Sudarshan–Lindblad (GKSL) equation~\cite{gorini1976completely, lindblad1976generators} for Markovian environments, while negative rates correspond to non-Markovian dynamics~\cite{Rivas:2014aa}. For $n$ qubits, direct numerical integration of \cref{eqn:time_local_ME} is limited to $n\approx 10$, as directly evolving the density matrix $\rho(t)$ has a memory cost that scales as $O(D^2)$, where $D=2^n$ is the Hilbert space dimension.

Several approximate methods address this scaling challenge. The quantum trajectory (QT) method~\cite{dalibard1992wave, gardiner1992wave, daley2014quantum} offers a quadratic memory scaling improvement (to $O(D)$) by unraveling system-bath interactions into stochastic processes, but struggles to converge when \cref{eqn:time_local_ME} violates complete positivity (CP) \cite{becker2023quantum}. Tensor-network based algorithms~\cite{vidal, verstraete2004matrix, zwolak2004mixed, werner2016positive} represent quantum states, operations, and noise channels as interconnected tensors that can be efficiently contracted to compute expectation values while maintaining favorable scaling, but their performance degrades significantly with growing entanglement, limiting their scalability with increasing circuit depth~\cite{pan2021simulating}.

\begin{figure}
    \centering
    \includegraphics[width=1.0\linewidth]{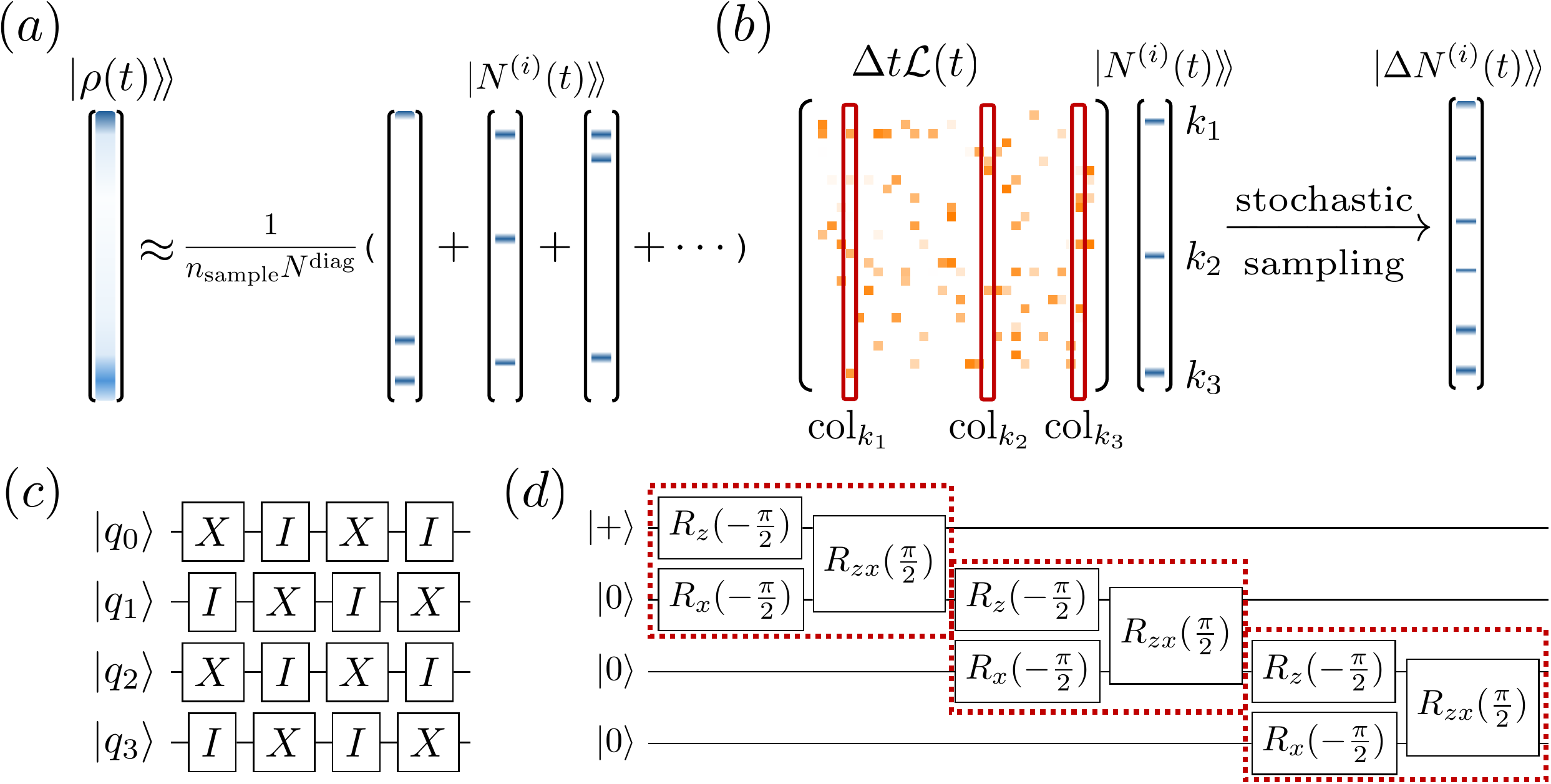}
    \caption{(a) Algorithm illustration: A pseudo-sparse, vectorized density matrix $\kket{\rho(t)}$ is stochastically compressed by averaging a set of integer-valued, sparse population vectors. (b) The action of a sparse Liouvillian superoperator $\mathcal{L}(t)$ on each population vector $\kket{N^{(i)}(t)}$ over a small time step, $\Delta t$, is conducted via random sampling, ensuring that only columns in $\mathcal{L}(t)$ corresponding to non-zero elements in $\kket{N^{(i)}(t)}$ (shown in red bars) are evaluated, and the resulting $\kket{\Delta N^{(i)}(t)}$ remains sparse. (c-d) The quantum circuits we simulated here: (c) crosstalk suppression of freely-evolving qubits using the staggered $X\!X$ dynamical decoupling sequence and (d) $n$-qubit GHZ-state preparation. Each dashed box implements a CNOT gate.}
    \label{fig:schematic_and_circuits}
\end{figure}

An open system density matrix $\rho(t)$ evolving via a quantum circuit tends to becomes pseudo-sparse in the computational basis. The primary reason is that off-diagonal elements typically decay rapidly due to relaxation and dephasing. A crucial consequence we exploit here is that the density matrix can be stochastically compressed across multiple copies [see \cref{fig:schematic_and_circuits}(a)] without introducing bias upon averaging. Moreover, even in the classically harder regime of transient dynamics, where the decoherence time exceeds the circuit duration, one can perform moderate over-truncation in the stochastic compression to confine the operational subspace dimension to $O(\lambda D)$ , with $\lambda \ll 1$. Compensating for over-truncation effects by averaging over multiple samples, the bias introduced remains minimal~\cite{supp}, and substantial scaling improvements can be achieved by evolving the compressed density matrix.

Building on these observations, here we introduce a Quantum Monte Carlo (QMC) method that stochastically compresses the density matrix using a finite number of walkers in the computational basis, discretely approximating it and emulating its time evolution through population dynamics. By tracking only occupied states while preserving ergodicity through walker sets averaging, this approach significantly reduces computational and memory requirements. The population dynamics here generalize those of full configuration interaction QMC (FCIQMC)~\cite{booth2009fermion} and its variants~\cite{blunt2014density, malone2015interaction, guther2018time, nagy2018driven, petras2021sign}, demonstrating effective control over the sign problem common in QMC methods, enabling efficient long-time simulations at large scales, and further handling cases where CP is violated. By exploiting the trace-preserving property of master equations, we not only eliminate the need for population control and its associated error~\cite{ghanem2021population} in FCIQMC but also ensure the ergodicity that permits the combination of multiple memory-light simulations, achieving statistically unbiased simulations that enable a qualitative leap in scalability while also, as shown in our benchmark examples, outperforming the state-of-the-art QT method implemented in QuTiP~\cite{Johansson:2012aa} with improved sampling~\cite{abdelhafez2019gradient} by several orders of magnitude.

\textit{QMC algorithm}.---We column-vectorize the $D\times D$-dimensional density matrix $\rho(t)$ in the computational basis $\{\ket{i}\}_{i=1}^{D}$, yielding the superket $\kket{\rho(t)} = \mathrm{vec}[\rho(t)]$, and rewrite \cref{eqn:time_local_ME} as $\frac{d}{dt}\kket{\rho(t)} = \mathcal{L}(t) \kket{\rho(t)}$, where $\mathcal{L}(t)$ is the matrix representing the Liouvillian superoperator whose form after vectorization~\cite{roth1952equations} is given by
\begin{align}
    \mathcal{L}(t) = &-i \mathbb{I} \otimes H(t) + i H^{\rm T}(t) \otimes \mathbb{I} \label{eqn:liouvillian_matrix} \\
    &+ \frac{1}{2}\sum_k \gamma_k(t) (2L_k \otimes L_k - \mathbb{I}\otimes L_k^{\dagger}L_k - L_k^{\dagger}L_k \otimes \mathbb{I}). \nonumber
\end{align}
While $\mathcal{L}(t)$ is  $D^2\!\times\!D^2$-dimensional, as shown in \cref{fig:schematic_and_circuits}(b), we only need to store the columns corresponding to non-zero elements of the compressed vector—each containing at most $O(n)$ nonzero elements~\cite{supp}. Consequently, for a QMC algorithm operating in a subspace of dimension $O(\lambda D)$, at most $O(n\lambda D)$ elements need to be stored in memory.

\begin{figure*}
    \centering
    \includegraphics[width=1.00\linewidth]{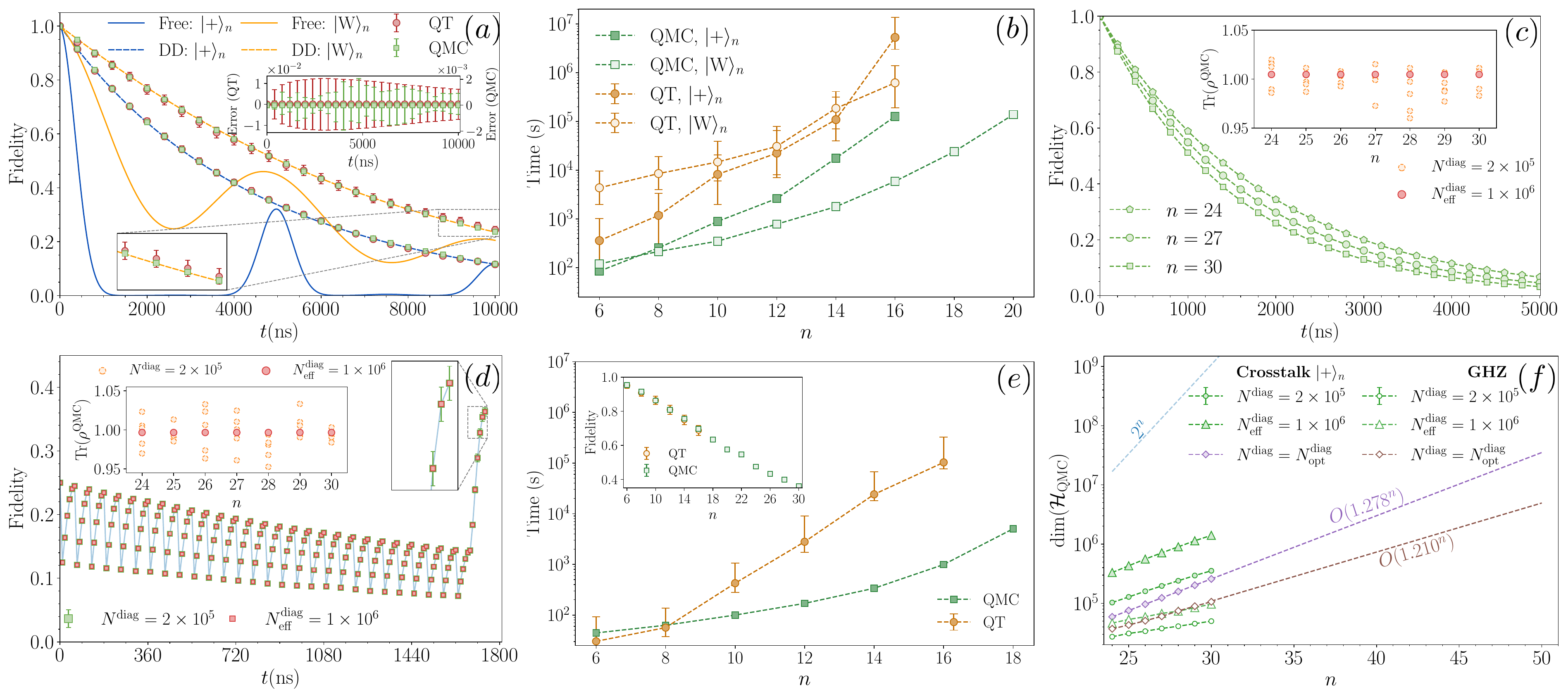}
    \caption{(a) Fidelity of the 10-qubit states $\ket{+}_{10}$ and $\ket{\rm W}_{10}$ under free evolution and DD over $10^4$ ns. Solid/dashed: exact master equation (free/DD). QT and QMC averages were computed with comparable total runtimes: for $\ket{+}_{10}$—3 QMC samples with $N^{\rm diag}=10^6$ (3793 s) vs 5800 QT trajectories (4013 s); for $\ket{\rm W}_{10}$—4 QMC samples with $N^{\rm diag}=5{\times}10^6$ (677 s) vs 1200 QT trajectories (765 s). Inset: magnified error comparison for $\ket{+}{10}$ (QT: left axis; QMC: right), with QMC errors $\approx$10× smaller. Free-evolution data match the exact solution and are omitted for clarity. (b) Computational runtime vs qubit number for QT and QMC under DD over $5{\times}10^3$ ns. (c) QMC results for $\ket{+}$ state evolution at $n=24,27,30$ under DD using replica aggregation: five replicas at $N^{\rm diag}=2{\times}10^5$ combined into $N_{\rm eff}^{\rm diag}=10^6$. Inset: ${\rm Tr}(\rho^{\rm QMC})$ for each replica and the aggregate over $n=24-30$. (d) QMC results for 30-qubit GHZ-state preparation dynamics; averages from five replicas ($N^{\rm diag}=2{\times}10^5$) and their aggregate. (e) Computational runtime for $n$-qubit GHZ-state preparation. Inset: final-state fidelity vs $n$; QT averaged over 5000 trajectories. QMC averaged over 4 samples with $N^{\rm diag}=10^6$ for $n=6-18$ and aggregated from 5 replicas with $N^{\rm diag}=2{\times}10^5$ for $n=22-30$. (f) Effective QMC subspace size vs $n$ for two circuits (crosstalk-suppressed $\ket{+}$ evolution and GHZ preparation), shown alongside $2^n$, for fixed $N^{\rm diag}=2{\times}10^5$, aggregated $N_{\rm eff}^{\rm diag}{=}10^6$. Fits indicate $O(1.210^n)$ and $O(1.278^n)$ scaling with optimized $N^{\rm diag}_{\rm opt}$.}
    \label{fig:MarkovianDynamicsBenchmark}
\end{figure*}

To stochastically compress the density matrix within the QMC framework, despite the sign problem~\cite{Troyer:2005aa} from both non-stoquastic Hamiltonians~\cite{Bravyi:QIC08,Bravyi:2009sp,Marvian:2019aa,Gupta_2020} and coherent evolution~\cite{church2021real}, we define a set of complex-signed walkers~\cite{booth2013towards}, $\{w^{(\alpha)}_{t} \}^{N^{\mathrm{tot}}_t}_{\alpha=1}$, where $N^{\mathrm{tot}}_t$ is the total number of walkers at time $t$, as a stochastic representation of $\kket{\rho(t)}$. 
The $\alpha$'th signed walker $w_t^{(\alpha)}$ is defined as $w_t^{(\alpha)} \equiv s_\alpha(t) \mathrm{loc}_\alpha$ with a complex-valued sign ${s}_\alpha(t) \in \{1, -1, +i, -i\}$ and where $\mathrm{loc}_\alpha=\kket{i,j} = \mathrm{vec}(\ket{i}\otimes\ket{j})$ denotes the location of the $\alpha$'th walker; $\kket{N(t)} = \sum_{\alpha=1}^{N^{\mathrm{tot}}_t} w^{(\alpha)}_{t}$ represents an integer-valued population vector. As illustrated in \cref{fig:schematic_and_circuits}(a), the link from the integer-valued population vectors of individual samples, $\kket{N^{(i)}(t)}$, to the physical, continuous-valued density matrix $\kket{\rho(t)}$, is then established through an ensemble average, up to normalization:
\begin{equation}
    \kket{\rho(t)}\approx \frac{1}{n_{\rm sample }N^{\rm diag}} \sum_{i}^{n_{\rm sample}} \kket{N^{(i)}(t)},
    \label{eqn:walker_averaging}
\end{equation}
where $N^{\rm diag}$ is the number of diagonal walkers, $N^{\rm diag} \equiv \sum_{\{\alpha| \mathrm{loc}_\alpha \in \{ \kket{i,i}\} \}}w^{(\alpha)}_{t}$. As the Liouvillian is trace-preserving, i.e., $\Tr\left[\mathcal{L}(t)\kket{\rho(t)}\right] = 0$, $N^{\rm diag}$ is a time-independent constant. To initialize the simulation, signed walkers are distributed to discretely represent a target population vector derived from the initial density matrix: $\kket{N(0)} := N^{\mathrm{diag}} \kket{\rho(0)}$. Consequently, $N^{\mathrm{diag}}$ sets both the target precision and the computational overhead: larger $N^{\mathrm{diag}}$ yields smaller statistical error and higher memory/runtime overhead.

Replacing $\kket{\rho(t)}$ with $\kket{N(t)}$ necessitates computing the incremental population vector, $\Delta t \,\mathcal{L}(t)\,\kket{N(t)}$. To keep this incremental vector sparse, we approximate matrix-vector multiplication by random sampling, implemented via a two-step spawn and annihilation population dynamics procedure~\cite{booth2009fermion, nagy2018driven} which we detail in the End Matter. Crucially, in the \emph{annihilation step}, the sign problem is addressed by explicit dynamic cancellations~\cite{spencer2012sign}. Operationally, as detailed in the End Matter, we suppress the sign problem by maintaining a sufficiently large $N^{\mathrm{diag}}$, which enables efficient annihilation of opposite-sign walkers and thereby supports trace preservation. The master equation is solved by evolving multiple $\kket{N(t)}$ using spawn-annihilation population dynamics, with their average providing an unbiased estimate for $\kket{\rho(t)}$. 

\textit{Markovian circuit dynamics}.---To demonstrate the accuracy and efficiency of our method, we simulate multiple paradigmatic circuits implemented on superconducting transmon qubits~\cite{clarke2008superconducting, kjaergaard2020superconducting}. The effective system Hamiltonian for uniform qubit frequency $\omega_q$ and uniform, undesired, always-on $Z\!Z$ coupling (crosstalk) strength $J$
can be written as
$H_S = - \omega_q \sum_{i=1}^{n} \sigma_i^{z} + J\sum_{\langle i,j \rangle} \sigma_i^{z} \sigma_j^{z}$. The system is subject to local Markovian amplitude damping and dephasing, with respective Lindblad operators $L_k = \sigma^{-}_{k}$ and $L_k = \sigma^{z}_{k}$, and corresponding $T_1 = 100 \ \mu\text{s}$ and $T_2 = 50 \ \mu\text{s}$. We set $\omega_q = 5$ GHz and $J = 100$ kHz, matching current superconducting transmon-qubit devices~\cite{transmon-invention}. Since $\omega_q \gg J$, we work in the rotating frame and apply the rotating-wave approximation, which eliminates the qubit frequency terms~\cite{tripathi2022suppression, ezzell2023dynamical}. This also allows us to implement quantum gates using square pulses, with durations of $10$ ns for single-qubit gates and $50$ ns for two-qubit gates, unless otherwise specified.

We first consider Ramsey-like experiments, where we measure the fidelity $\mathcal{F} = |\bra{\psi_0} \rho_t \ket{\psi_0}|$ between an initial state $\ket{\psi_0}$ and the freely evolved state $\rho_t$, and use dynamical decoupling (DD) to suppress the crosstalk between neighboring qubits~\cite{tripathi2022suppression}. As shown in \cref{fig:schematic_and_circuits}(b), we adopt a staggered $X\!X$ sequence, suitable for crosstalk suppression in the multi-qubit setting~\cite{Jones_1999,Zhou2023,Shirizly:2024aa,niu2024multi,evert2024syncopated,brown2024efficient}: for evenly indexed qubits we apply the sequence $f_{\tau/2} - X - f_{\tau} - X - f_{\tau/2}$, while for oddly indexed qubits the sequence is shifted by $\tau/2$, yielding $f_{\tau} - X - f_{\tau} - X$, where $\tau$ is the pulse interval, or idle time.
We perform simulations with two different initial $n$-qubit states: the maximally coherent product state $\ket{+}_n$ and the entangled W-state, $\ket{\rm W}_n$, $\frac{1}{\sqrt{n}}(\ket{100\dots0} + \ket{010\dots0} + \dots + \ket{000\dots1})$. In the computational (Pauli $Z$) basis, the $\ket{+}_n$ state's density matrix is fully dense—requiring walkers on all $4^n$ states at initialization—so we transform to the Pauli $X$ basis where the initial state is sparse.
In this basis, amplitude damping increases coherence and decreases sparsity over time, presenting the most challenging case for QMC. In \cref{fig:MarkovianDynamicsBenchmark}(a), we benchmark a 10-qubit system, the largest size where numerically exact solutions of \cref{eqn:time_local_ME} are available. We compare the QT and QMC results subject to DD using similar computation times. As shown in the inset, QMC exhibits significantly smaller error bars despite using fewer statistical samples for both initial states. This demonstrates QMC's improved convergence rate.

To estimate the relative computational costs of QMC and QT at equivalent precision levels, we fix the QMC error bar and determine the range of quantum trajectories needed to match the precision. We compare 4 QMC samples with 5000 QT trajectories across increasing system sizes until QT reaches memory limits. For each system size and measurement, we calculate the ratio of QT to QMC error bars. Since QT trajectories are independent, for $N_{\mathrm{traj}}$ trajectories the error bar scales as $O(N_{\mathrm{traj}}^{-1/2})$. Thus, the estimated QT computational time is the measured time for 5000 trajectories multiplied by the squared error bar ratio. \cref{fig:MarkovianDynamicsBenchmark}(b) presents the median estimated QT runtime, with error bars representing the range between the minimum and maximum estimates. For the least favorable case for QMC (the $\ket{+}_n$ state), QMC achieves more than a tenfold speedup compared to QT's best-case runtime, clearly demonstrating QMC's scaling advantage over QT for any initial state under similar experimental conditions. For the sparser $\ket{\rm W}_n$ state, QMC outperforms QT by approximately $100\times$ as QT is unable to leverage pseudo-sparsity.

Another advantage unique to QMC is that, provided the evolution of $\kket{N(t)}$ is trace-preserving, independent replicas can be aggregated without increasing per-replica memory overhead:
\begin{equation}
    \kket{\rho^{\rm QMC}(t)} = \frac{1}{n_{\rm sample}N^{\rm diag}_{\rm eff}} \sum_i^{n_{\rm sample}} \sum_{j}^{r} \kket{N^{(i,j)}(t)}.
\end{equation}
Here $r$ memory-light replicas with per-replica walker counts $N^{\rm diag}$ are combined into an unbiased estimator with $N^{\rm diag}_{\rm eff} = r N^{\rm diag}$, increasing the effective sample size linearly (error $\propto (N^{\rm diag}_{\rm eff})^{-1/2}$) and thereby extending simulatable system size without raising per-replica RAM. To illustrate this, we aggregate five independent replicas for crosstalk-suppressed $|+\rangle_n$ evolution: each replica uses $N^{\rm diag}=2\times 10^5$ walkers, yielding an unbiased estimator with $N^{\rm diag}_{\rm eff}=10^6$. \cref{fig:MarkovianDynamicsBenchmark}(c) reports fidelities for $n=24$, $27$, and $30$. The trace is then estimated via diagonal walker normalization as $\Tr(\rho^{\rm QMC}_j) = N^{\rm diag}_j(t_f) / N^{\rm diag}$ for a single replica and $\Tr(\rho^{\rm QMC}_{\rm eff}) = \sum_j^5 N^{\rm diag}_j(t_f) / N^{\rm diag}_{\rm eff}$ for the aggregate, where $N^{\rm diag}_j(t_f)$ is the measured diagonal walker counts for the $j$'th replica at the end of the circuit. The inset shows per-replica estimates as points slightly scattered around unity, while the aggregate contributes a single value closer to $1$, reflecting increased precision across $n=24-30$.

Next, we consider GHZ state preparation experiments. As depicted in \cref{fig:schematic_and_circuits}(d), the circuit is initialized in the state $\ket{+0\cdots 0}$, followed by a sequence of CNOT gates. Similar agreement of QMC with exact solutions, along with smaller QMC error bars at fixed runtime than QT, are provided in the Supplement \cite{supp}. In \cref{fig:MarkovianDynamicsBenchmark}(d), we simulate a 30-qubit GHZ state preparation circuit with QMC using the same replica-aggregation scheme and plot the fidelity to the ideal GHZ state. Each peak, appearing every 60 ns, marks a full CNOT gate cycle. The final bump in the last 60 ns corresponds to the last CNOT and indicates that the GHZ state is essentially prepared. In \cref{fig:MarkovianDynamicsBenchmark}(e), using the same protocol, QMC demonstrates significantly better efficiency and scaling—outperforming QT by a factor of $100$ at larger sizes—and, with replica aggregation, extends the simulation to 30 qubits, whereas QT is memory-constrained to 16. Overall, these results highlight QMC's improved scaling, precision, and efficiency over QT across quantum circuits with varying entanglement characteristics, especially for large-scale systems.

As a quantitative scaling metric, we define the effective QMC subspace size $\dim(\mathcal{H}_{\text{QMC}}) := \bigl\vert \{ \text{loc} \,\mid\, \text{loc} = \text{loc}_\alpha \text{ for some } \alpha \in \{1,\dots,N_t^{\rm tot}\}\}\bigr\vert = O(\lambda D)$, and the optimized per-replica walker number at system size $n$ and tolerance $\tau$ as $N^{\rm diag}_{\rm opt}(n,\tau) := \min\{N^{\rm diag}\in \mathbb{N}: \epsilon(n, N^{\rm diag})\leq \tau \}$, where $\epsilon(n, N^{\rm diag})$ is our metric for Hermiticity preservation (see End Matter). Using the simulated five replicas $N^{\rm diag}=2\times10^5$ ,we form aggregates with $N^{\rm diag}_{\rm eff}\in\{4,6,8,10\}\times10^5$ and fit the empirical model
$\dim(\mathcal{H}_{\text{QMC}}) = Ce^{\beta n} (N^{\rm diag})^{\gamma}$ to our data, obtaining $(\beta,\gamma)=(0.208, 0.721)$ for crosstalk-suppressed $|+\rangle_n$ evolutions and $(\beta,\gamma)=(0.130, 0.425)$ for GHZ circuits. From this fit we compute $N^{\rm diag}_{\rm opt}(n,\tau)$ at $\tau=0.02$ and then simulate with the optimized walker counts to extract scaling. The resulting $\dim(\mathcal{H}_{\text{QMC}})$ curves are shown in \cref{fig:MarkovianDynamicsBenchmark}(f), from which we read off the effective $n$-scaling and project to $n\approx 50$. Given that storing and sampling $\dim(\mathcal{H}_{\text{QMC}})=4\times10^6$ requires 64 GB RAM, we estimate per-replica memory for 50-qubit runs of $\sim560$ GB (DD $|+\rangle_n$) and $\sim 80$ GB (GHZ). The inferred sparsity lies in the range $\lambda = 10^{-8} -10^{-10}$. This makes simulations at larger $n$ practical with moderate computational resources, well before exponential growth becomes the bottleneck.

\textit{Non-Markovian dynamics}.---The GKSL equation, despite its widespread use in open quantum systems, is strictly valid only in the Markovian regime~\cite{breuer2002theory}. Including non-Markovian effects may violate complete positivity, causing QT methods to fail to converge even at short times due to negative jump probabilities~\cite{rivas_open_2012}. QMC, however, maintains convergence to exact solutions even in the non-Markovian setting as it directly emulates the master equation and, as we have shown, suppresses the sign problem through walker annihilation. This makes it even more advantageous: beyond its favorable scaling, it can be applied to broad classes of open systems.

To demonstrate QMC's convergence advantage relative to QT in the non-Markovian setting, we consider two qubits collectively coupled to a shared bosonic bath. The model Hamiltonian is
    $H = \sum_{i=1}^2 \omega_i \sigma^{+}_i\sigma^{-}_i + \sum_k \epsilon_k b_k^\dagger b_k +\sum_k g_k b_k S^+ +\mathrm{h.c.}$
Here,  $S^+ = \sigma^{+}_1 +\sigma^{+}_2$, $b_k$ are bosonic lowering operators, $\omega_j$ and $\epsilon_k$  denote the characteristic frequencies of the qubits and the bath modes, and  $g_k$  are coupling constants. We perform the Born-Markov approximation~\cite{breuer2002theory} and diagonalize the rate matrix to arrive at the Redfield master equation~\cite{mozgunov2020completely}:
\begin{equation}
    \frac{d}{dt}\rho(t) = -i \sum_{i,j=1}^2 A_{ij} [\sigma^+_{j}\sigma^-_{i},\rho(t)] + \sum_{k}^2 \lambda_k \mathcal{D}[L_k] \rho(t),
    \label{eqn:NonMarkovME}
\end{equation}
where the dissipators $\mathcal{D}[L_k]$ are linear combinations of $\sigma^+$ and $\sigma^-$ (see End Matter). The rates are given by $\lambda_k = \frac{\gamma_1 + \gamma_2}{4} \pm \sqrt{\frac{\gamma_1^2 + \gamma_2^2 + 8\kappa^2}{8}}$, where $\gamma_j = 2\int_{-\infty}^{\infty} dt \sum_{k}|g_k|^2 e^{i(\omega_j - \epsilon_k)t}$, and can clearly be negative, a signature of non-Markovianity~\cite{rivas2010entanglement}. We rotate to the orthonormal basis defined by the Lindblad operators: $\ket{1} = L_1^{\dagger} \ket{0}, \ket{2} = L_2^{\dagger} \ket{0}, \ket{3} = L_2^{\dagger} L_1^{\dagger} \ket{0}$ and then simulate \cref{eqn:NonMarkovME} by tracking the population of the density matrix elements in the new basis, $\tilde{\rho}_{00}$, $\tilde{\rho}_{11}$ and $\tilde{\rho}_{22}$ with the initial state $\frac{1}{\sqrt{3}}(\ket{0} + \ket{1} + \ket{2})$. There exist different flavors of QT for handling the negative jump probabilities~\cite{breuer1999stochastic,hush2015generic,becker2023quantum}; we adopt QuTiP's~\cite{Johansson:2012aa} default method that uses the influence martingale technique~\cite{donvil2022quantum}.

\begin{figure}
    \centering
    \includegraphics[width=1.0\linewidth]{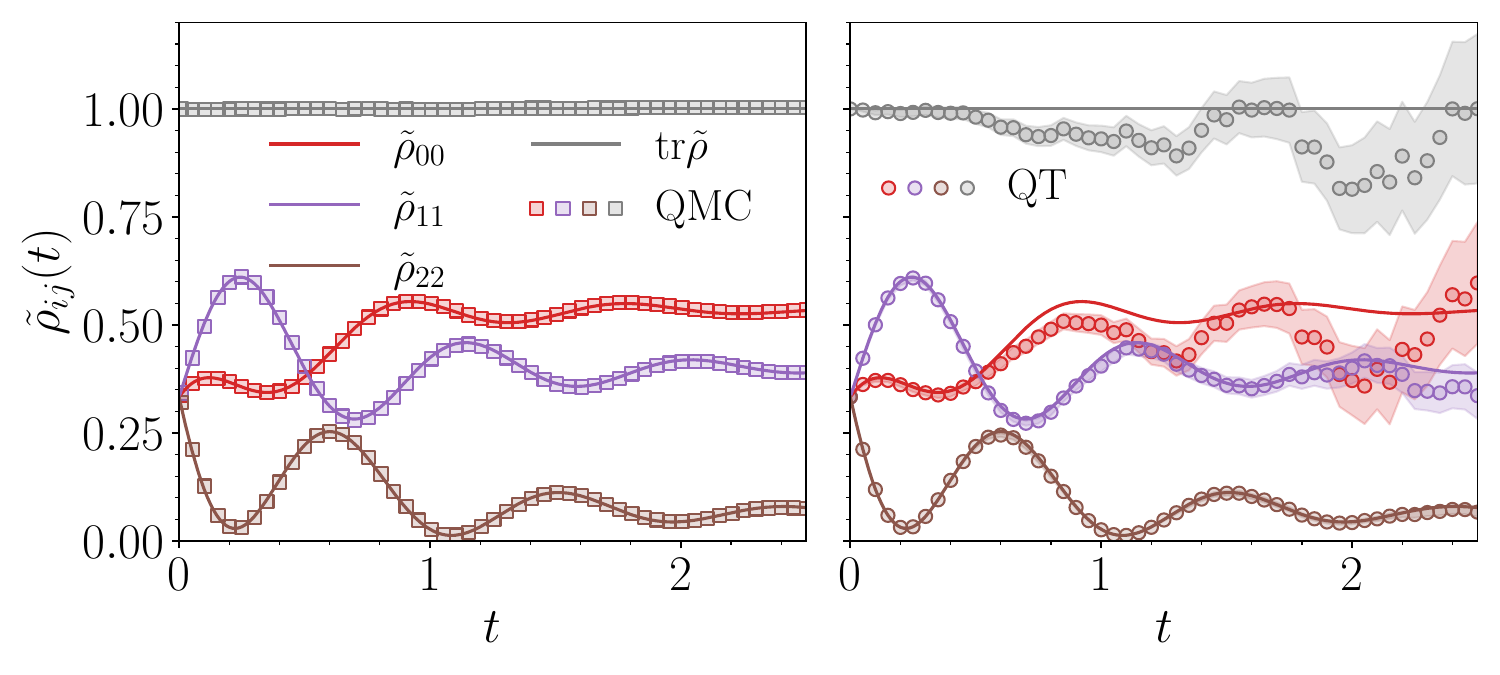}
    \caption{Non-Markovian dynamics for density matrix elements $\tilde{\rho}_{00}$, $\tilde{\rho}_{11}$ and $\tilde{\rho}_{22}$. Solid lines represent exact master equation solutions in both panels. Left: Results from a single QMC sample with $10^6$ walkers; state traces are calculated as the ratio $\frac{N^{\rm diag}(t)}{N^{\rm diag}(0)}$. Right: QT solutions averaged over $10^4$ trajectories, with state traces computed from averaged influence martingale values. Parameter values: $\omega_1=0.25$, $\omega_2 = 0.5$, $\gamma_1 = 1$, $\gamma_2 = 4$, $\alpha=3$, $\kappa=1$, yielding $\lambda_1 = -0.5178$ and $\lambda_2 = 3.0178$. See~\cite{supp} for similar results for the off-diagonal elements.}
    \label{fig:NonMarkov}
\end{figure}

In \cref{fig:NonMarkov}, we compare the convergence of QMC and QT with exact solutions. QT is accurate only for short times, with errors (shown as shaded areas) increasing uncontrollably at later times. QT's average trace, calculated by averaging influence martingale values per trajectory, quickly deviates from 1 during the evolution, indicating failed convergence. In contrast, QMC exhibits perfect agreement for both ground and excited states while preserving the trace throughout the entire simulation using just a single sample of $10^6$ walkers, demonstrating its effectiveness in the non-Markovian setting. 

\textit{Conclusion.}---%
By stochastically compressing and evolving the density matrix, dynamically suppressing the sign problem and using statistically optimized sampling, the QMC algorithm we have presented here achieves significant computational speedups and superior system-size scaling over QT methods. This positions QMC as a more efficient and scalable classical simulator for simulating quantum error-correction and various quantum algorithms evolving under realistic noise models than any existing alternative, with far-reaching implications not only for quantum hardware and algorithm development but also for addressing master-equation-based problems in other fields, such as dissipative phase transitions~\cite{diehl2008quantum}, chemical reactions and transport phenomena~\cite{wichterich2007modeling, esposito2010self}, and quantum biology~\cite{palmieri2009lindblad, lambert2013quantum, Mohseni2014Quantum}.

\textit{Acknowledgments}.---%
The authors acknowledge the Center for Advanced Research Computing (CARC) at the University of Southern California for providing computing resources. This material is based upon work supported by, or in part by, the U. S. Army Research Laboratory and the U.S. Army Research Office under contract/grant number W911NF2310255.

\textit{End Matter on spawn and annihilation population dynamics}.---%
In the \emph{spawn step}, a state $\kket{i,j}$ occupied by $N_{ij}(t)$ walkers with sign $s_{ij}(t)$ at time $t$ can spawn new walkers into any state $\kket{k,l}$ satisfying $\Re(\mathcal{L}_{ij}^{kl}(t))\neq 0$ or $\Im(\mathcal{L}_{ij}^{kl}(t))\neq 0$. Here, $\mathcal{L}_{ij}^{kl}(t) = \bbra{k,l} \mathcal{L}(t) \kket{i,j}$ is the $(i,j;k,l)$ element of the Liouvillian matrix defined in \cref{eqn:liouvillian_matrix}. The number of newly spawned walkers, $N^{\mathrm{sp}}_{ij}(t)$, is drawn from a binomial distribution, $N^{\mathrm{sp}}_{ij}(t) \sim B(N^{\mathrm{sp}}_{ij}(t), p^{\mathrm{sp}}_{ij}(t))$, where $p^{\mathrm{sp}}_{ij}(t) \equiv (\sum_{k,l} |\Re (\mathcal{L}_{ij}^{kl}(t))| + |\Im (\mathcal{L}_{ij}^{kl}(t))|)\Delta t$. Those $N^{\mathrm{sp}}_{ij}(t)$ walkers are then distributed across the real and imaginary parts of all connected states via a multinomial distribution:
    $p_{ij\rightarrow (kl,c)} = \frac{|c (\mathcal{L}_{ij}^{kl}(t))|}{\sum_{k'l'} | \Re(\mathcal{L}_{ij}^{k'l'}(t))| + | \Im(\mathcal{L}_{ij}^{k'l'}(t))|},$
where $c \in \{\Re , \Im\}$. Because each spawned walker must carry a consistent sign, we assign it based on whether the real or imaginary component of $\mathcal{L}_{ij}^{kl}(t)$ is chosen. For the real channel, the sign is $s_{ij}(t) \mathrm{sgn}( \Re(\mathcal{L}_{ij}^{kl}(t))$, while for the imaginary channel it is $i s_{ij}(t) \mathrm{sgn}( \Im(\mathcal{L}_{ij}^{kl}))$. Consequently, the overall spawning step is unbiased~\cite{supp}, satisfying
$\mathbb{E}[\sum_{\alpha}^{N_t^{\mathrm{sp}}} w_{t+\Delta t}^{(\alpha)} \bigm| \kket{N(t)}] = \Delta t \mathcal{L} \kket{N(t)}$,
where $N_t^{\mathrm{sp}}=\sum_{ij} N^{\mathrm{sp}}_{ij}(t)$.

In the \emph{annihilation step}, the sign problem is addressed by explicit dynamic cancellations~\cite{spencer2012sign}. The newly spawned walkers, $\{w_{t+\Delta t}^{(\alpha')}\}_{\alpha'}^{N_t^{\mathrm{sp}}}$, merge with $\{w_{t}^{(\alpha)}\}_{\alpha}^{N^{\mathrm{tot}}_t}$, producing the population vector at time $t+\Delta t$: $\kket{N(t + \Delta t)} = \sum_{\alpha=1}^{N^{\mathrm{tot}}_t} w^{(\alpha)}_{t} + \sum_{\alpha'=1}^{N_t^{\mathrm{sp}}} w^{(\alpha')}_{t+\Delta t}$, with any walkers in the same state but carrying opposite signs ($\pm 1$ or $\pm i$) annihilating each other. Unlike other QMC algorithms that perform sign cancellation only at simulation's end, this method manages the sign problem dynamically. This approach prevents the accumulation of sign errors over time, making it, to our knowledge, the only current QMC algorithm capable of simulating noisy quantum circuits with time-dependent Hamiltonians, with one notable exception tailored for the single-site impurity model~\cite{cohen2013numerically}.

\textit{End matter on simulation bias and ergodicity}.---%
The simulation is unbiased and ergodic in that for a single sample ($n_{\rm sample} = 1$), the statistical error, $\epsilon_t \equiv\kket{\rho(t)} - \kket{N(t)}$, asymptotically approaches $0$ as~\cite{supp}
\begin{equation}
    \mathbb{E}\left[\Vert \epsilon_t \Vert_2^2 \mid \kket{N(t)}\right] \leq \Lambda(t) \frac{2 N_t^{\mathrm{tot}}}{(N^{\rm diag})^2} ,
    \label{eqn:error_bound}
\end{equation}
with rate $O(\frac{N_t^{\mathrm{tot}}}{(N^{\rm diag})^2}) \approx O(\frac{1}{N^{\rm diag}})$. Here, $\Lambda(t) = \max_{1\leq \alpha \leq N_t^{\mathrm{tot}}} (\Delta t)^2\Vert \mathrm{col}_\alpha[\mathcal{L}(t)] \Vert_0 \Vert \mathrm{col}_\alpha[\mathcal{L}(t)] \Vert_2^2 + \frac{1}{4}$, where, for the $\alpha$th walker with $\mathrm{loc}_\alpha = \kket{i,j}$, $\mathrm{col}_\alpha[\mathcal{L}(t)]$ denotes the $(i,j)$th column of $\mathcal{L}(t)$. Note that since $\mathrm{loc}_\alpha$ indicates only occupied states, unoccupied states do not affect the bound. While not tight, this bound suggests that even a non-sparse system (with relatively large $\Vert\mathrm{col}_\alpha[\mathcal{L}(t)] \Vert_0$) can still be simulated to a given error using moderate computational resources (quantified by $N_t^{\mathrm{tot}}$), provided the elements of $\Delta t\mathcal{L}(t)$ are small enough to keep $\Lambda(t)$ sufficiently small. In~\cite{supp}, we show that the matrix form of $\kket{N(t)}$ approaches a positive semidefinite matrix as $N^{\rm diag}$ increases, indicating that a single control parameter, $N^{\rm diag}$, can be used to balance both the approximation error $\| \epsilon_t\|_2$ and the deviation from positivity against computational cost. 

\textit{End Matter on simulation methodology}.---%
\cref{eqn:walker_averaging} enables us to evolve multiple sparse vectors, $\kket{N^{(i)}(t)}$, rather than a dense $\kket{\rho(t)}$. Although the first-order Euler solver,
$
\kket{\rho(t + \Delta t)} \approx \kket{\rho(t)} + \Delta t \mathcal{L}(t) \kket{\rho(t)},
$
is commonly used in the QMC literature for imaginary-time, closed-system evolution, we find that it introduces larger time-step errors in real-time, open-system scenarios. We therefore adopt the more accurate second-order Adams–Bashforth (AB2) solver,
$
    \kket{\rho(t + 2\Delta t)} \approx \kket{\rho(t + \Delta t)} + \Delta t [\frac{3}{2}\mathcal{L}(t+\Delta t) \kket{\rho(t + \Delta t)} -\frac{1}{2}\mathcal{L}(t) \kket{\rho(t)}]
$
to mitigate these errors at the expense of doubling the memory and time cost. 

To ensure a fair comparison between QT and QMC, both methods utilized the AB2 solver, with QT using slightly larger adaptive time steps compared to QMC's fixed time steps. All simulations were performed on identical hardware: 16 cores of an AMD EPYC 7542 CPU with 64 GB RAM. For QT, parallelization was implemented across all trajectories, while for QMC, parallelization was applied across all walker-occupied states~\cite{nagy2018driven}, and samples were processed sequentially. The QMC computational time reflects the cumulative time for all sequential samples. The error bars indicate 95\% confidence intervals computed using bootstrapping. All codes \cite{src}, scripts and data \cite{data_repo} needed to reproduce the results in this paper are available online.

Lastly, the initiator approximation from FCIQMC~\cite{cleland2010communications,ghanem2019unbiasing} can be applied to our method, where we discard spawned walkers to unoccupied states if they originate from an occupied state with walker population below a threshold $\xi N^{\rm diag}$. This truncation introduces only negligible errors that can be extrapolated to zero as $\xi \rightarrow 0$. In our simulations, we set $\xi=0.1\%$ and observe substantial speedup without compromising precision.

\textit{End Matter on Hermiticity preservation}.---For systems beyond exactly solvable sizes, we use the Frobenius-norm anti-Hermiticity $\epsilon(n, N^{\rm diag}) := \lVert \rho^{\rm QMC} - (\rho^{\rm QMC})^{\dagger} \lVert_F$ as a numerically tractable internal-consistency diagnostic, since a direct positivity test is infeasible. 
Using the same simulated five replicas, we fit the empirical models $\epsilon(n, N^{\rm diag}) = Ae^{\alpha n} (N^{\rm diag})^{-1/2} + \epsilon_{\infty}$ to our data \cite{supp}, obtaining $\alpha=0.0262$ for crosstalk-suppressed $|+\rangle_n$ evolutions and $\alpha=0.0715$ for GHZ circuits.

\begin{figure}[t]
    \centering
    \includegraphics[width=1.0\linewidth]{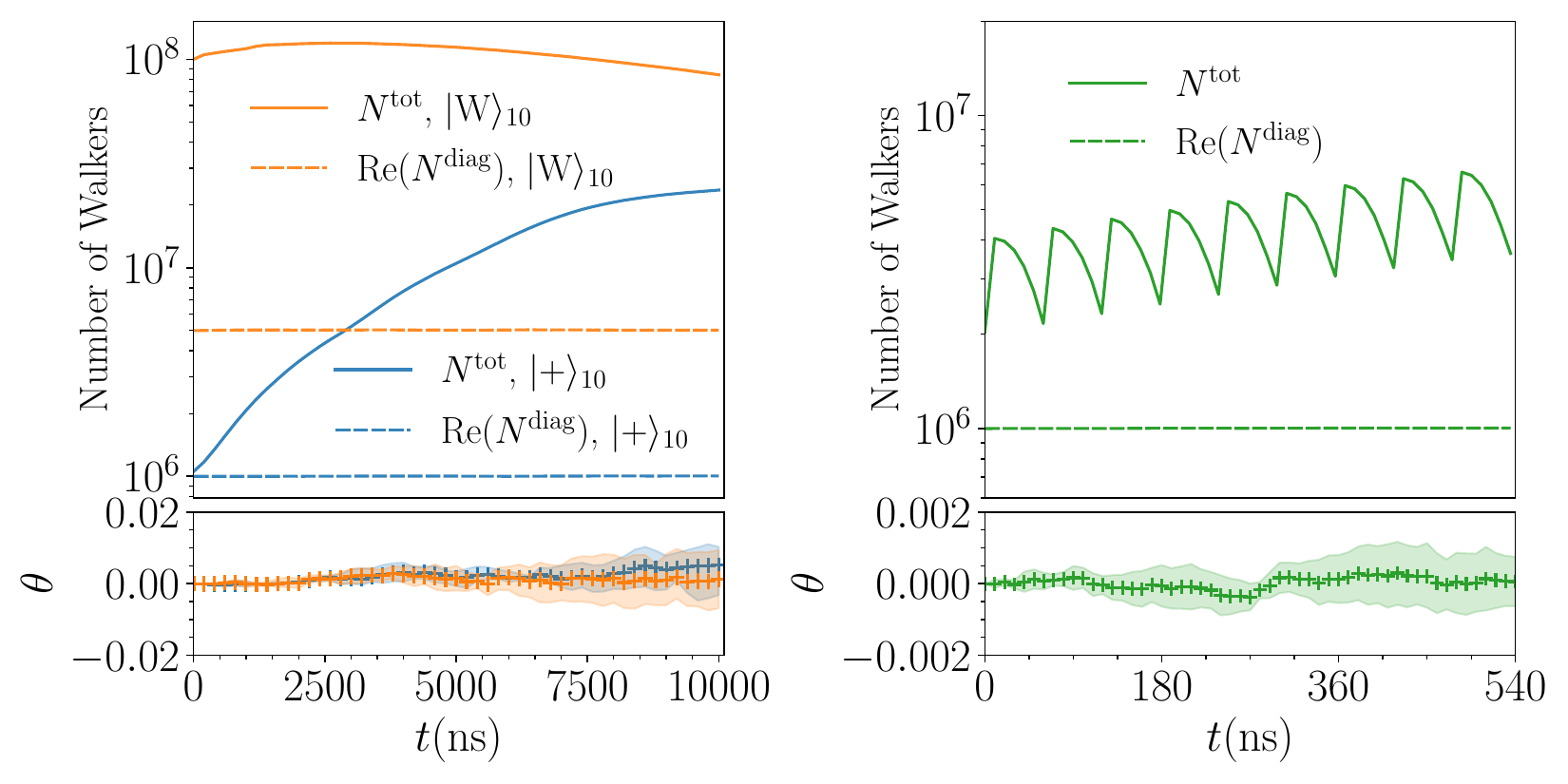}
    \caption{Walker population dynamics for two 10-qubit simulations: crosstalk suppression (left) and GHZ state preparation (right). The real components of diagonal populations remain consistently stable across all cases during evolution, with negligible error bars. Bottom panels show phase angles of diagonal populations, demonstrating dynamical sign suppression.}
    \label{fig:WalkerDynamics}
\end{figure}

\textit{End Matter on sign-problem suppression}.---Building on our scaling analysis, the diagnostics in \cref{fig:WalkerDynamics} validate unbiased cross-replica aggregation. Although $\mathcal{L}(t)$ is trace-preserving, stochastically evolving $\kket{N(t)}$ does not by itself keep $N^{\rm diag}$ fixed; if $N^{\rm diag}$ drifts, the QMC-approximated density matrix becomes unphysical and sample averaging breaks down. Since a physical density matrix is Hermitian, any imaginary component of $N^{\rm diag}$ arises from the sign problem, so we monitor $\theta := \arctan\frac{{\rm Im}(N^{\rm diag})}{{\rm Re}(N^{\rm diag})}$ to quantify it. \Cref{fig:WalkerDynamics} plots $N^{\mathrm{tot}}_t$, ${\rm Re}(N^{\rm diag})$, and $\theta$ for the $\ket{+}_{10}$, $\ket{\mathrm{W}}_{10}$, and $\ket{\mathrm{GHZ}}_{10}$ simulations. With sufficiently large initial walker numbers, ${\rm Re}(N^{\rm diag})$ stays essentially constant—eliminating the need for heuristic population control techniques and their associated biases—and the phase angles remain below $0.02$, indicating effective annihilation-based sign problem suppression. Because $N^{\mathrm{tot}}_t$ counts off-diagonal walkers, it tracks coherence: it grows for $\ket{+}_{10}$ in the Pauli $X$ basis (noise-induced coherence), decreases for $\ket{\mathrm{W}}_{10}$ under decoherence, and increases during GHZ preparation as CNOT gates entangle the system. These trends provide a practical diagnostic for when to allocate more walkers to maintain accurate, trace-preserving, sign-problem-suppressed QMC. Additional results supporting these conclusions are presented in Ref.~\cite{supp}.

\textit{End Matter on the Redfield equation}.---The frequency matrix given in \cref{eqn:NonMarkovME} is
\begin{equation}
    A = \begin{pmatrix}
    \omega_1 + \alpha & \alpha + \frac{\kappa}{2} - i\frac{\gamma_1 -\gamma_2}{8} \\
    \alpha + \frac{\kappa}{2} - i\frac{\gamma_2 -\gamma_1}{8} & \omega_2 + \alpha + \kappa
    \end{pmatrix} .
\end{equation}
The decoherence rates $\gamma_j$ are given in the main text; the remaining terms are the Lamb shift and are given by: 
\begin{align}
    \alpha &= \Im\int_{0}^{\infty} dt \sum_{k}|g_k|^2 e^{i(\omega_1 - \epsilon_k)t} \label{eqn:lamb_shift_1} \\
    \kappa &= \Im \int_{0}^{\infty} dt \sum_{k}|g_k|^2 \left(e^{i(\omega_2 - \epsilon_k)t} - e^{i(\omega_1 - \epsilon_k)t}\right). \label{eqn:lamb_shift_2}
\end{align}

Despite QMC's success in this example, potential limitations arise for large systems, where diagonalizing the full system Hamiltonian to obtain the Redfield equation [\cref{eqn:NonMarkovME}] becomes a major computational bottleneck. However, recent work~\cite{schnell2023global} demonstrates that an accurate approximation of the memory kernel can be achieved without explicit diagonalization, opening new possibilities for applying our QMC formalism to these approximated Redfield equations at larger scales.



\section*{Supplementary material}

# Supplemental Material: Real-time Quantum Monte Carlo Algorithm For Noisy Quantum Circuit Simulations

Tong Shen[1,2] and Daniel A. Lidar[1,2,3,4]

[1]*Department of Electrical and Computer Engineering,*
*University of Southern California, Los Angeles, California 90089, USA*
[2]*Center for Quantum Information Science & Technology,*
*University of Southern California, Los Angeles, California 90089, USA*
[3]*Department of Chemistry, University of Southern California, Los Angeles, California 90089, USA*
[4]*Department of Physics and Astronomy, University of Southern California, Los Angeles, California 90089, USA*

We maintain the same notation established in the main text throughout the Supplemental Material:

| Symbol | Description |
|---|---|
| $n$ | Number of qubits in the system |
| $D$ | Dimension of the Hilbert space, $2^n$ |
| $s_\alpha(t)$ | Sign of the $\alpha$th walker at time $t$, $s_k(t) \in \{1, -1, +i, -i\}$ |
| $\text{loc}_\alpha$ | Location of the $\alpha$th walker, $\text{loc}_\alpha \in \{\|\|i,j\rangle\rangle\|\}_{i,j=1}^{D}$ |
| $w_t^{(\alpha)}$ | A single walker with index $k$ at time $t$, $w_t^{(\alpha)} := s_\alpha(t)\|i,j\rangle\rangle$ with $\text{loc}_\alpha = \|i,j\rangle\rangle$ |
| $N_t^{\text{tot}}$ | Total number of walkers at time $t$ |
| $N^{\text{diag}}$ | Total number of diagonal walkers |
| $N_{ij}(t)$ | Number of walkers in state $\|i,j\rangle\rangle$, always a positive integer |
| $\|N(t)\rangle\rangle$ | Population vector at time $t$, $\|N(t)\rangle\rangle := \sum_{i=1}^{N_t^{\text{tot}}} w_t^{(k)}$ |
| $\|\rho(t)\rangle\rangle$ | Vectorized density matrix at time $t$ |
| $\mathcal{L}(t)$ | Liouvillian matrix |
| $\text{col}_\alpha[\mathcal{L}(t)]$ | $(i,j)$th column vector of the Liouvillian matrix corresponding to the $\alpha$th walker with $\text{loc}_\alpha = \|i,j\rangle\rangle$ |

## S1. ALGORITHMIC DETAILS

In this section, we detail the population dynamics algorithm implemented in our QMC method. This algorithm is inspired by the work of Booth et al. [1], who developed it for time-independent, wavefunction-based quantum chemistry calculations. The initial framework only employed real-valued walkers for spawning and annihilation and was later extended to complex-valued walkers by Guther et al. [2] to handle real-time evolutions. A significant advancement was made by Blunt et al. [3], who generalized the method to density-matrix-based walker dynamics, enabling mixed state simulations. Building on this formalism, Nagy et al. [4] demonstrated how walkers can be evolved using Liouvillian operators, allowing simulations of open systems in non-equilibrium steady states. These progressive developments established the algorithmic framework necessary for our extension to the full-time dynamics of open systems.

Essentially, the population dynamics emulate the real-time propagation

$$\begin{aligned}
|\Delta\rho(t)\rangle\rangle &\approx \Delta t \mathcal{L}(t)|\rho(t)\rangle\rangle \\
&= \Delta t \mathcal{L}(t) \sum_{ij} \langle\langle i,j|\rho(t)\rangle\rangle |i,j\rangle\rangle \\
&= \sum_{kl}\sum_{ij} \langle\langle i,j|\rho(t)\rangle\rangle \Delta t \mathcal{L}_{ij}^{kl}(t)|k,l\rangle\rangle
\end{aligned} \qquad (S1)$$

where the first approximation comes from the Euler solver for ordinary differential equations, and generalizations to higher-order solvers that are constructed as linear combinations of Euler steps follow automatically. Notice that by definition, $\langle\langle i,j|\rho(t)\rangle\rangle$ relates to the number of walkers in state $|i,j\rangle\rangle$, $N_{ij}(t)$, via

$$N_{ij}(t) \approx s_{ij}(t) N^{\text{diag}} \langle\langle i,j|\rho(t)\rangle\rangle, \qquad (S2)$$

with $N^{\text{diag}}$ assumed sufficiently large to maintain trace preservation, i.e., $N^{\text{diag}}$ remain nearly constant, and $s_{ij}(t)$ is the sign of the state, $s_{ij}(t) \in \{1, -1, +i, -i\}$. By averaging over samples, we obtain:

$$\langle\langle i,j|\rho(t)\rangle\rangle = \frac{1}{N^{\text{diag}}} s_{ij}(t) \mathbb{E}[N_{ij}(t)]. \tag{S3}$$

It follows that the incremental population in state $|k,l\rangle\rangle$ is

$$\Delta N_{kl}(t) \approx N^{\text{diag}} \langle\langle k,l|\Delta\rho(t)\rangle\rangle = N^{\text{diag}} \sum_{ij} \langle\langle i,j|\rho(t)\rangle\rangle \Delta t \mathcal{L}_{ij}^{kl}(t) \approx \sum_{ij} \Delta t \mathcal{L}_{ij}^{kl}(t) s_{ij}(t) N_{ij}(t), \tag{S4}$$

indicating that new walkers in state $|k,l\rangle\rangle$ are walkers spawned from all the connected states for which $|\mathcal{L}_{ij}^{kl}(t)| > 0$. Here, $\Delta N_{kl}(t)$ is generally a complex-valued random variable, and on average

$$\begin{aligned}
\mathbb{E}[\Delta N_{kl}(t) \mid |N(t)\rangle\rangle] &= \sum_{ij} \Delta t \mathcal{L}_{ij}^{kl}(t) s_{ij}(t) N_{ij}(t) \\
&= \sum_{ij} \Delta t \left[ s_{\mathcal{L}_{ij}^{kl}(t)}^{\text{Re}} |\text{Re}(\mathcal{L}_{ij}^{kl}(t))| + s_{\mathcal{L}_{ij}^{kl}(t)}^{\text{Im}} |\text{Im}(\mathcal{L}_{ij}^{kl}(t))| \right] s_{ij}(t) N_{ij}(t),
\end{aligned} \tag{S5}$$

where for a general, complex-valued Liouvillian matrix element, we use $s_{\mathcal{L}_{ij}^{kl}(t)}^{\text{Re}} \in \{+1, -1\}$ and $s_{\mathcal{L}_{ij}^{kl}(t)}^{\text{Im}} \in \{+i, -i\}$ to denote the signs of its real and imaginary part, respectively. Thus, a stochastic sampling that fulfills Eq. (S5) in expectation provides an unbiased estimate of Eq. (S1). This sampling can be implemented as either a walker-based scheme—common in the conventional FCIQMC literature [1]—or a state-based scheme [4], which is more efficient and is used in the main text.

For the walker-based sampling, a walker in state $|i,j\rangle\rangle$ first randomly selects a connected state $|k,l\rangle\rangle$ with uniform probability $p_{ij \to (kl,c)}^{\text{gen}} = 1/n_{\text{conn}}$, where $c \in \{\text{Re}, \text{Im}\}$ and $n_{\text{conn}}$ denotes the total number of states connected to $|i,j\rangle\rangle$, i.e., $n_{\text{conn}} = \|\text{col}_\alpha[\mathcal{L}(t)]\|_0$. A spawning event then occurs at a rate

$$p_{ij \to (kl,c)}^{\text{spawn}} = \frac{\Delta t |c(\mathcal{L}_{ij}^{kl}(t))|}{p_{ij \to (kl,c)}^{\text{gen}}} = \Delta t |c(\mathcal{L}_{ij}^{kl}(t))| n_{\text{conn}}. \tag{S6}$$

If the spawning event is successful, the newly created walker is assigned a sign $s_{\mathcal{L}_{ij}^{kl}(t)}^c s_{ij}(t)$. On average, the total number of newly spawned walkers in $|k,l\rangle\rangle$ originating from $|i,j\rangle\rangle$ is

$$\mathbb{E}[\Delta N_{kl}(i,j;t) \mid N_{ij}(t)] = \sum_c N_{ij}(t) p_{ij \to (kl,c)}^{\text{spawn}} p_{ij \to (kl,c)}^{\text{gen}} = \sum_c s_{\mathcal{L}_{ij}^{kl}(t)}^c \Delta t |c(\mathcal{L}_{ij}^{kl}(t))| s_{ij}(t) N_{ij}(t). \tag{S7}$$

Summing over $(i,j)$ the recovers Eq. (S5).

In the walker-based scheme, each walker in state $|i,j\rangle\rangle$ at time $t$ is treated independently, despite having the same sign and spawning rules. This redundancy reduces efficiency, since all such walkers can be collectively treated as a single unit. Consequently, instead of iterating over individual walkers at each time step, one can iterate over the currently occupied states in a statistically equivalent way. In the state-based scheme, as presented in the main text, we treat the entire population $N_{ij}(t)$ at once. First, we determine the number of newly spawned walkers through binomial sampling:

$$N_{ij}^{\text{sp}}(t) \sim B(N_{ij}(t), p_{ij}^{\text{sp}}(t)), \tag{S8}$$

where

$$p_{ij}^{\text{sp}}(t) := \Delta t \left( \sum_{kl} |\text{Re}(\mathcal{L}_{ij}^{kl}(t))| + |\text{Im}(\mathcal{L}_{ij}^{kl}(t))| \right). \tag{S9}$$

On average, we have $\mathbb{E}[N_{ij}^{\text{sp}}(t) \mid N_{ij}(t)] = p_{ij}^{\text{sp}}(t) N_{ij}(t)$. Next, as shown in the main text, these spawned walkers are distributed among all connected states according to a multinomial distribution:

$$p_{ij \to (kl,c)} = \frac{|c(\mathcal{L}_{ij}^{kl}(t))|}{\sum_{k'l'} |\text{Re}(\mathcal{L}_{ij}^{k'l'}(t))| + |\text{Im}(\mathcal{L}_{ij}^{k'l'}(t))|}. \tag{S10}$$

and again, each spawned walker's sign is $s^c_{\mathcal{L}^{kl}_{ij}(t)} s_{ij}(t)$. The expectation value of newly spawned walkers in $|k,l\rangle\rangle$ originating from $|i,j\rangle\rangle$ is then

$$\mathbb{E}[\Delta N_{kl}(i,j;t) \mid N_{ij}(t)] = \sum_c p_{ij\to(kl,c)} \mathbb{E}[N^{\text{sp}}_{ij}(t) \mid N_{ij}(t)]$$
$$= \sum_c p_{ij\to(kl,c)} p^{\text{sp}}_{ij}(t) N_{ij}(t)$$
$$= \sum_c s^c_{\mathcal{L}^{kl}_{ij}(t)} \Delta t |c(\mathcal{L}^{kl}_{ij}(t))| s_{ij}(t) N_{ij}(t), \tag{S11}$$

which matches Eq. (S7) and shows that both schemes are statistically equivalent. However, the state-based approach handles the much smaller set of occupied states, compensating for the added complexity of multinomial sampling Eq. (S10) compared to the simpler binomial sampling Eq. (S6).

Since $\Delta N_{kl}(t) = \sum_{\alpha'} w^{(\alpha')}_{t+\Delta t}$ with $\text{loc}_{\alpha'} = |k,l\rangle\rangle$, summing over $(k,l)$ on both sides of Eq. (S5) yields

$$\mathbb{E}[\sum_{kl} \Delta N_{kl}(t) \mid |N(t)\rangle\rangle] = \mathbb{E}[\sum_{\alpha'}^{N^{\text{sp}}_t} w^{(\alpha')}_{t+\Delta t} \mid |N(t)\rangle\rangle]$$
$$= \sum_{kl}\sum_{ij} \Delta t \left[ s^{\text{Re}}_{\mathcal{L}^{kl}_{ij}(t)} |\text{Re}(\mathcal{L}^{kl}_{ij}(t))| + s^{\text{Im}}_{\mathcal{L}^{kl}_{ij}(t)} |\text{Im}(\mathcal{L}^{kl}_{ij}(t))| \right] s_{ij}(t) N_{ij}(t)$$
$$= \sum_{kl}\sum_{ij} \Delta t \mathcal{L}^{kl}_{ij} s_{ij}(t) N_{ij}(t)$$
$$= \Delta t \mathcal{L} |N(t)\rangle\rangle \tag{S12}$$

Here, $N^{\text{sp}}_t := \sum_{ij} N^{\text{sp}}_{ij}(t)$. This proves that the spawning step yields an unbiased estimate of Eq. (S1).

In the annihilation step, the spawned walkers from all currently occupied states, $\{w^{(\alpha')}_{t+\Delta t}\}^{N^{\text{sp}}_t}_{\alpha'}$, are merged with the current walker set, $\{w^{(k)}_t\}^{N^{\text{tot}}_t}_k$. Pairs in the same state with opposite signs ($\pm 1$ or $\pm i$) annihilate each other and yield the updated walker set for the next time step, $\{w^{(\alpha)}_{t+\Delta t}\}^{N^{\text{tot}}_{t+\Delta t}}_\alpha$. Note that due to annihilation, the total number of walkers does not follow a simple additive relation: $N^{\text{tot}}_{t+\Delta t} \neq N^{\text{sp}}_t + N^{\text{tot}}_t$.

The complete process is outlined in Algorithm 1, where in our implementation the innermost loop is parallelized across available CPU cores. Although the pseudo-code provides a general measurement procedure, the observable $O$ is typically required to be local for efficiency, such as the fidelity measurements in the main text or the expectation values of stabilizer operators in quantum error correction codes. Measuring a global observable effectively "uncompresses" $|N(t)\rangle\rangle$, thereby losing the memory savings offered by the stochastic compression.

## S2. ERROR ANALYSIS

In this section, we prove that for a single sample ($n_{\text{sample}} = 1$), increasing the number of walkers eventually converges the QMC results to the exact solution at the rate $O(1/N^{\text{diag}})$. This confirms that the QMC algorithm is ergodic, meaning a single sample with a sufficient number of walkers can emulate the exact evolution. However, since error bars cannot be obtained from a single sample, we opt to use multiple samples. This proof extends the FCIQMC convergence analysis [5] in two ways: from closed, time-independent systems to open, time-dependent ones, and from iterating each real-valued walker individually to iterating over states occupied by complex-valued walkers.

Without loss of generality, we focus on the Euler solver,

$$|N(t+\Delta t)\rangle\rangle \approx \left[\Delta t \mathcal{L}(t) + \mathbb{I}\right] |N(t)\rangle\rangle, \tag{S13}$$

and treat the stochastic sampling of the population vector under the Liouvillian supermatrix as an inexact matrix-vector multiplication. The resulting convergence proof naturally extends to multi-step linear solvers, including the second-order Adams–Bashforth (AB2) solver used in the main text, as such methods can be expressed as linear combinations of Euler solvers. The incremental vector $|\Delta N(t)\rangle\rangle$ at time $t$ can be viewed as the exact evolution plus an error term:

$$|\Delta N(t)\rangle\rangle = \sum_{\alpha=1}^{N^{\text{tot}}_t} w^{(\alpha)}_{t+\Delta t} = \Delta t \mathcal{L}(t) |N(t)\rangle\rangle + N^{\text{diag}} \epsilon_t, \tag{S14}$$

**Algorithm 1:** Real-time Open QMC Algorithm

**Initialization:** Set the diagonal walker number $N^{\text{diag}}$ and construct the initial population vector based on the given initial state:
$$|N(0)\rangle\!\rangle := N^{\text{diag}}|\rho(0)\rangle\!\rangle.$$

**for** $n = 1 : n_{\text{sample}}$ **do**
    **for** $t = 0 : \Delta t : t_f$ **do**
        Initialize an empty set for spawned walkers: $\{w_{t+\Delta t}^{(\alpha)}\}_\alpha$.
        **for** *each occupied state* $\text{loc}_k = |i,j\rangle\!\rangle$ *with* $N_{ij}(t)$ *walkers* **do**
            **Spawning:**
            Identify all states $|k,l\rangle\!\rangle$ connected to $|i,j\rangle\!\rangle$ and compute $p_{ij}^{\text{sp}}(t)$ as in Eq. (S9).
            Use binomial sampler with sample size $N_{ij}(t)$ and success rate $p_{ij}^{\text{sp}}(t)$:
$$N_{ij}^{\text{sp}}(t) \leftarrow B(N_{ij}(t), p_{ij}^{\text{sp}}(t))$$
            Fill the spawned walker set with the multinomial distribution defined in Eq. (S10):
$$\{w_{t+\Delta t}^{(\alpha)}\}_\alpha \leftarrow \text{Multinomial}(N_{ij}^{\text{sp}}(t), p_{ij\rightarrow(kl,c)})$$
            Assign proper signs to the spawned walkers
        **end**
        **Annihilation:** Merge the spawned walkers into the population vector:
$$|N(t+\Delta t)\rangle\!\rangle \leftarrow |N(t)\rangle\!\rangle + \sum_\alpha w_{t+\Delta t}^{(\alpha)},$$
        where pairs of walkers at the same location with opposite signs ($\pm 1$ or $\pm i$) are annihilated and removed. The diagonal walker population is updated using the spawned walkers occupying the diagonal elements of the density matrix:
$$N_{t+\Delta t}^{\text{diag}} \leftarrow N_t^{\text{diag}} + \left\|\text{Re}\left(\text{diag}(\{w_{t+\Delta t}^{(\alpha)}\})\right)\right\|_0 + i \cdot \left\|\text{Im}\left(\text{diag}(\{w_{t+\Delta t}^{(\alpha)}\})\right)\right\|_0.$$
        **Measurement:** If a measurement is required at $t + \Delta t$, compute the expectation value for the observable $O$ with the estimator:
$$\text{tr}\left(O \frac{1}{N_{t+\Delta t}^{\text{diag}}}|N(t+\Delta t)\rangle\!\rangle\right).$$
    **end**
**end**
**Averaging:** Average expectation value results across different samples using bootstrapping.

---

where $\epsilon_t$ is introduced to denote the normalized error arising from the inexact matrix-vector multiplication. Since $|\Delta N(t)\rangle\!\rangle = \sum_{ij} \Delta N_{ij}(t)$, we know from Eq. (S12) that the approximation is unbiased, i.e.,

$$\mathbb{E}\left[\Delta \vec{N}(t+\Delta t) \mid |N(t)\rangle\!\rangle\right] = \Delta t \mathcal{L}(t)|N(t)\rangle\!\rangle, \tag{S15}$$

and

$$\mathbb{E}[\epsilon_t \mid |N(t)\rangle\!\rangle] = 0. \tag{S16}$$

However, this does not show how the error decreases as we increase the number of walkers, and correspondingly the computational cost. Therefore, we examine the variance of the error ($L^2$ norm) conditioned on the previous time step, $\mathbb{E}[\|\epsilon_t\|_2^2 \mid |N(t)\rangle\!\rangle]$. Since each walker evolves independently,

$$\Delta t \mathcal{L}(t)|N(t)\rangle\!\rangle = \Delta t \mathcal{L}(t) \cdot \sum_{\alpha=1}^{N_t^{\text{tot}}} w_t^{(\alpha)} = \sum_{\alpha=1}^{N_t^{\text{tot}}} \Delta t \mathcal{L}(t) w_t^{(\alpha)}, \tag{S17}$$

$\Delta t \mathcal{L}(t) w_t^{(\alpha)}$ and $\Delta t \mathcal{L}(t) w_t^{(\alpha')}$ are independent for $\alpha \neq \alpha'$. Therefore, by denoting $\Xi_t := N^{\text{diag}} \epsilon_t$, the error term can

be decomposed as errors from individual walkers:

$$\mathbb{E}[\|\Xi_t\|_2^2 \mid |N(t)\rangle\!\rangle] = \mathbb{E}\left[\|\sum_{\alpha=1}^{N_t^{\text{tot}}}\left(w_{t+\Delta t}^{(\alpha)} - \Delta t \mathcal{L}(t) w_t^{(\alpha)}\right)\|_2^2 \mid |N(t)\rangle\!\rangle\right] \tag{S18a}$$

$$= \sum_{\alpha=1}^{N_t^{\text{tot}}} \mathbb{E}\left[\left(w_{t+\Delta t}^{(\alpha)} - \Delta t \mathcal{L}(t) w_t^{(\alpha)}\right)^\top \cdot \left(w_{t+\Delta t}^{(\alpha)} - \Delta t \mathcal{L}(t) w_t^{(\alpha)}\right) \mid |N(t)\rangle\!\rangle\right] \tag{S18b}$$

$$+ 2 \sum_{1 \leq \alpha < \alpha' < N_t^{\text{tot}}} \left(\mathbb{E}\left[w_{t+\Delta t}^{(\alpha)} - \Delta t \mathcal{L}(t) w_t^{(\alpha)} \mid |N(t)\rangle\!\rangle\right] \cdot \mathbb{E}\left[w_{t+\Delta t}^{(\alpha')} - \Delta t \mathcal{L}(t) w_t^{(\alpha')} \mid |N(t)\rangle\!\rangle\right]\right) \tag{S18c}$$

$$= \sum_{\alpha=1}^{N_t^{\text{tot}}} \mathbb{E}\left[\|w_{t+\Delta t}^{(\alpha)} - \Delta t \mathcal{L}(t) w_t^{(\alpha)}\|_2^2 \mid |N(t)\rangle\!\rangle\right], \tag{S18d}$$

where Eq. (S18c) vanishes due to Eqs. (S14) and (S16).

Therefore, it suffices to derive an error bound for a single walker. Here, $w_{t+\Delta t}^{(\alpha)}$ corresponds to the walker spawned from $w_t^{(\alpha)}$ and can potentially be 0 if the spawning event is unsuccessful. Without loss of generality, we consider the $\alpha$'th walker with $\text{loc}_\alpha = |i,j\rangle\!\rangle$, and perform the analysis with walker-based sampling (which is statistically equivalent to the state-based scheme). The spawned walker $w_{t+\Delta t}^{(\alpha)}$ is then distributed as

$$w_{t+\Delta t}^{(\alpha)} = \begin{cases} s_\alpha(t) s^c_{\mathcal{L}_{ij}^{kl}(t)} |k,l\rangle\!\rangle & \text{w.p. } \frac{|c(\Delta t \mathcal{L}_{ij}^{kl}(t))|}{n_{\text{conn}}} \\ 0 & \text{w.p. } \frac{1 - |c(\Delta t \mathcal{L}_{ij}^{kl}(t))|}{n_{\text{conn}}} \end{cases} \tag{S19}$$

Here, $c \in \{\text{Re}, \text{Im}\}$ as defined in the main text. Note that

$$\Delta t \mathcal{L}(t) w_t^{(\alpha)} = s_\alpha(t) \sum_{c \in \{\text{Re}, \text{Im}\}} \sum_{kl} s^c_{\mathcal{L}_{ij}^{kl}(t)} |c(\Delta t \mathcal{L}_{ij}^{kl}(t))| \cdot |k,l\rangle\!\rangle, \tag{S20}$$

hence the sign factors do not affect the variance of $w_{t+\Delta t}^{(\alpha)} - \Delta t \mathcal{L}(t) w_t^{(\alpha)}$. Since $w_{t+\Delta t}^{(\alpha)}$ follows a generalized Bernoulli (categorical) distribution, we arrive at

$$\mathbb{E}[\|w_{t+\Delta t}^{(\alpha)} - \Delta t \mathcal{L}(t) w_t^{(\alpha)}\|_2^2 \mid |N(t)\rangle\!\rangle] = n_{\text{conn}} \sum_{kl} \sum_{c \in \{\text{Re}, \text{Im}\}} |c(\Delta t \mathcal{L}_{ij}^{kl}(t))|^2 + \frac{1}{n_{\text{conn}}} \sum_{kl} \sum_{c \in \{\text{Re}, \text{Im}\}} (1 - |c(\Delta t \mathcal{L}_{ij}^{kl}(t))|)|c(\Delta t \mathcal{L}_{ij}^{kl}(t))|$$

$$\leq n_{\text{conn}} \sum_{kl} \sum_{c \in \{\text{Re}, \text{Im}\}} |c(\Delta t \mathcal{L}_{ij}^{kl}(t))|^2 + \frac{1}{n_{\text{conn}}} \cdot 2 n_{\text{conn}} \cdot \frac{1}{4}$$

$$= \|\text{col}_\alpha[\mathcal{L}(t)]\|_0 \sum_{kl} \sum_{c \in \{\text{Re}, \text{Im}\}} |c(\Delta t \mathcal{L}_{ij}^{kl}(t))|^2 + \frac{1}{2}$$

$$\leq 2(\Delta t)^2 \|\text{col}_\alpha[\mathcal{L}(t)]\|_0 \|\text{col}_\alpha[\mathcal{L}(t)]\|_2^2 + \frac{1}{2}. \tag{S21}$$

In the first line, the first term represents the diagonal variance, and the second term corresponds to the off-diagonal covariance. The second line follows the fact that $(1 - |c(\Delta t \mathcal{L}_{ij}^{kl}(t))|)|c(\Delta t \mathcal{L}_{ij}^{kl}(t))|$ is in any case bounded by $\frac{1}{4}$. The final inequality holds because

$$\sum_{kl} \sum_{c \in \{\text{Re}, \text{Im}\}} |c(\mathcal{L}_{ij}^{kl}(t))|^2 = \|\text{col}_\alpha[\text{Re}(\mathcal{L}(t))]\|_2^2 + \|\text{col}_\alpha[\text{Im}(\mathcal{L}(t))]\|_2^2 \leq 2\|\text{col}_\alpha[\mathcal{L}(t)]\|_2^2.$$

Thus, summing over all walkers yields

$$\mathbb{E}[\|\Xi_t\|_2^2 \mid |N(t)\rangle\!\rangle] \leq 2 N_t^{\text{tot}} (\Delta t)^2 (\max_{1 \leq \alpha \leq N_t^{\text{tot}}} \|\text{col}_\alpha[\mathcal{L}(t)]\|_0 \|\text{col}_\alpha[\mathcal{L}(t)]\|_2^2 + \frac{1}{4}). \tag{S22}$$

Since $\Xi_t = N^{\text{diag}} \epsilon_t$, we can rewrite the above estimate as

$$\mathbb{E}[\|\epsilon_t\|_2^2 \mid |N(t)\rangle\!\rangle] \leq \frac{2 N_t^{\text{tot}}}{(N^{\text{diag}})^2} (\Delta t)^2 (\max_{1 \leq \alpha \leq N_t^{\text{tot}}} \|\text{col}_\alpha[\mathcal{L}(t)]\|_0 \|\text{col}_\alpha[\mathcal{L}(t)]\|_2^2 + \frac{1}{4}) = \frac{2 N_t^{\text{tot}}}{(N^{\text{diag}})^2} \Lambda(t). \tag{S23}$$

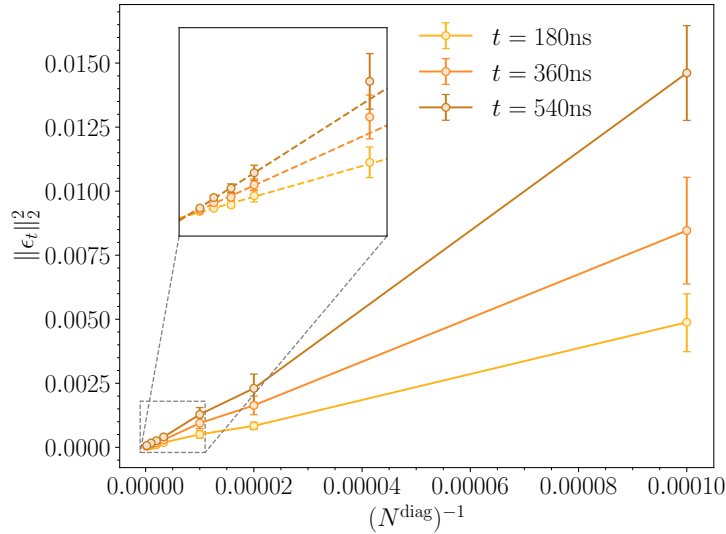

FIG. S1. Squared $L^2$-norm of the QMC error, $\mathbb{E}[\|\epsilon_t\|_2^2]$, averaged over 10 samples, in 10-qubit GHZ state preparation at 180 ns, 360 ns, and 540 ns, plotted as a function of the inverse diagonal walker number. The inset shows a linear fitting of the last four data points, with $r^2$ values of 0.9694, 0.9902, and 0.9972 for 180 ns, 360 ns, and 540 ns, respectively.

Assuming that $N^{\text{diag}}$ and $N_t^{\text{tot}}$ are of the same order, and that $\Lambda(t)$ can be upper-bounded by some constant $C$ that is much smaller than the walker number, i.e., $C \ll N^{\text{diag}}$, we can conclude that the error asymptotically converges to zero at a rate of $O(1/N^{\text{diag}})$.

In Fig. S1, we plot $\mathbb{E}[\|\epsilon_t\|_2^2]$ as a function of $(N^{\text{diag}})^{-1}$ to numerically verify the analytical scaling, using the 10-qubit GHZ state preparation experiment from the main text at three time points: 180 ns, 360 ns, and 540 ns. The QMC errors increase at later times due to error accumulation, but uniformly decrease as the diagonal walker number increases. In the inset, we perform linear fits on the last four data points and find $r^2$ values sufficiently close to 1, confirming the analytical scaling. The fifth data point deviates for the larger $t$ values from the linear fit because the linear error convergence rate in Eq. (S23) requires $N^{\text{diag}}$ to be a time-independent constant, which is violated when $N^{\text{diag}}$ is not sufficiently large. This issue is discussed further in the next section.

## S3. MORE ON WALKER DYNAMICS, SIGN PROBLEM AND POSITIVITY OF QMC DENSITY MATRIX

In this section, we examine how different initial diagonal walker numbers affect the behavior of QMC simulations, including convergence, accuracy, and the sign problem. We again use the 10-qubit GHZ state preparation experiment as the benchmarking example, as its relatively small system size enables the computation of various metrics essential for analysis, and its circuit, which introduces entanglement during evolution, exhibits more complex phenomena than circuits containing only single-qubit gates.

Trace preservation is not strictly maintained when matrix-vector multiplication is approximated by stochastic sampling, potentially leading to unphysical density matrix estimates and unreliable results. From the convergence analysis in the previous section, we can infer that increasing the walker number helps stabilize fluctuations in the trace (diagonal walker number). In Fig. S2(a), we plot the ratio of the diagonal walker number at time $t$ to its initial value, $\frac{N^{\text{diag}}(t)}{N^{\text{diag}}}$, with initial walker numbers of $10^3$, $10^4$ and $10^5$. When too few walkers are used (e.g., $N^{\text{diag}} = 10^3$), trace preservation breaks down at later times, deviating significantly from unity. Increasing the initial walker number to $10^5$ greatly suppresses these deviations, ensuring that subsequent measurements remain physically meaningful. Similarly to the main text, we use the phase angle $\theta = \arctan \frac{\text{Im}(N^{\text{diag}})}{\text{Re}(N^{\text{diag}})}$ to quantify the sign problem under different conditions. As shown in Fig. S2(b), the findings align with those of the trace preservation results: when too few walkers are initialized, a large number of unwanted imaginary walkers are spawned and remain on the diagonal. Initializing more walkers significantly reduces this unphysical accumulation of imaginary walkers.

Tracking the diagonal walker number and phase angle provides a convenient way to assess the robustness of QMC

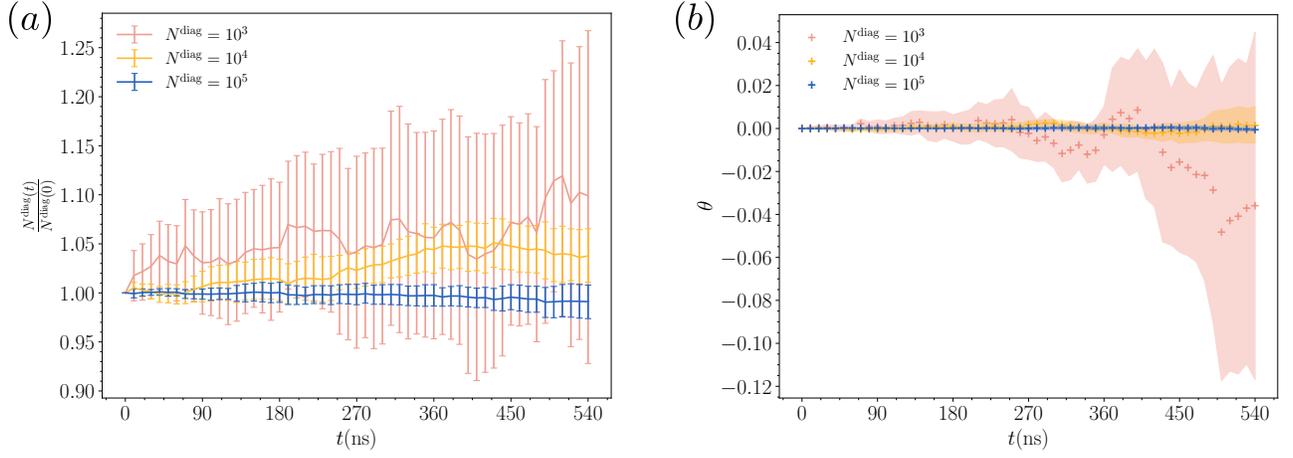

FIG. S2. (a) Change in diagonal walker number as a function of time in the 10-qubit GHZ state simulations, with initial walker numbers of $10^3$, $10^4$ and $10^5$. (b) Corresponding phase angle dynamics for each initial walker number. All results are averaged over 10 samples with bootstrapping.

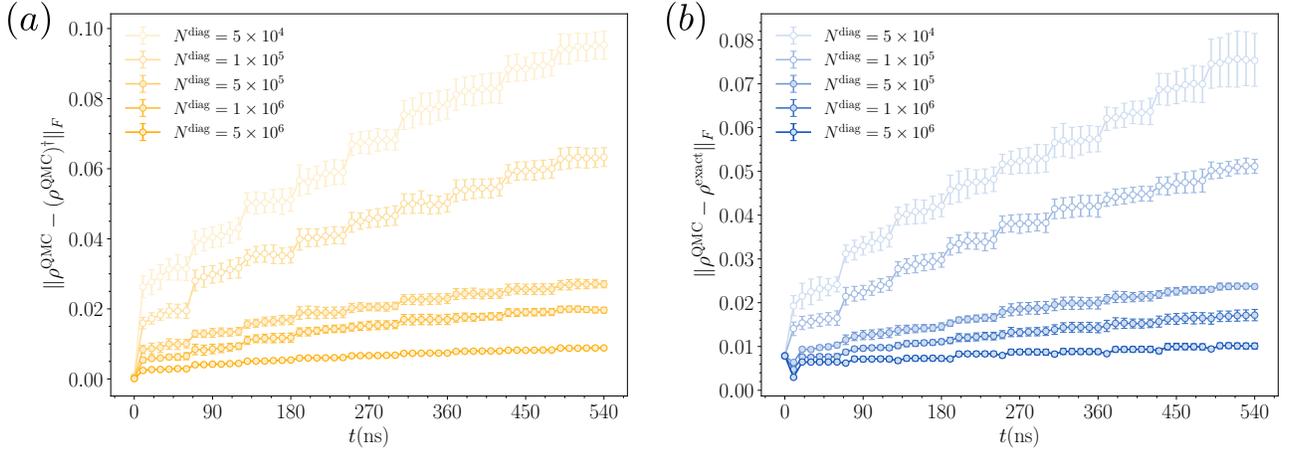

FIG. S3. (a) Time evolution of the closeness of the QMC-estimated density matrix to a Hermitian matrix, quantified using the Frobenius norm. (b) Time evolution of the Frobenius norm of the difference between the QMC-estimated density matrix and the exact density matrix. All results are averaged over 10 samples with bootstrapping.

simulations with minimal cost, and as shown in the main text, passing these checks corresponds to statistically unbiased and accurate measurements of local observables. However, merely monitoring the diagonal behavior does not offer enough insight into coherence and entanglement, as they reside in the off-diagonal elements of the density matrix expressed in the computational basis. Moreover, while the trace-preserving property of the CPTP map is ensured by initializing a sufficiently large number of walkers, the QMC algorithm itself does not explicitly enforce complete positivity. Consequently, a more rigorous robustness check is to verify that the QMC-approximated density matrix remains Hermitian and positive semidefinite.

In Fig. S3(a), we perform a Hermiticity check, measuring how closely the QMC-approximated density matrix $\rho^{\mathrm{QMC}}(t)$ matches its Hermitian conjugate, $(\rho^{\mathrm{QMC}}(t))^\dagger$. Here, $\rho^{\mathrm{QMC}}(t)$ is obtained by unvectorizing the average of 10 QMC samples, $\frac{1}{N^{\mathrm{diag}}} \sum_{i=1}^{10} |N^{(i)}(t)\rangle\!\rangle$. We quantify the deviation from Hermiticity via the Frobenius norm of the difference,

$$\|\rho^{\mathrm{QMC}}(t) - (\rho^{\mathrm{QMC}}(t))^\dagger\|_F,$$

plotted as a function of circuit time. For smaller walker numbers, $\rho^{\mathrm{QMC}}(t)$ not only deviates more strongly from Hermiticity but also exhibits larger sample-to-sample fluctuations (as shown by the sizable error bars). Increasing $N^{\mathrm{diag}}$ to $5 \times 10^6$ substantially reduces this deviation, keeping the Frobenius norm close to zero and growing very

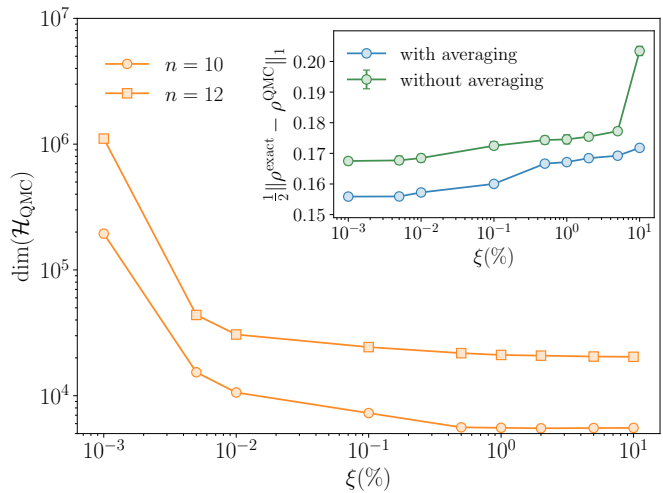

FIG. S4. $\dim(\mathcal{H}_{\text{QMC}})$ as a function of truncation ratio ($\xi$) for 10-qubit and 12-qubit GHZ state simulations. The inset shows the trace-norm distance from the exact solution for the 10-qubit case: green circles represent the average trace distance over five individual runs, while blue circles are the trace distance of the average over those same runs.

slowly with circuit time. This implies that time-accumulated stochastic errors have minimal impact on Hermiticity when the walker population is sufficiently large.

Positivity checks are more subtle. Although $\rho^{\text{QMC}}(t)$ may be close to Hermitian, it is never strictly Hermitian at any point in time, so its eigenvalues are generally complex, therefore $\rho^{\text{QMC}}(t)$ is not positive. Formally, one could construct a Hermitian matrix

$$\frac{\rho^{\text{QMC}}(t) + (\rho^{\text{QMC}}(t))^\dagger}{2}$$

and check whether all of its eigenvalues are positive. However, this procedure is computationally expensive. Instead, we use a simpler measure: we evaluate the Frobenius norm of the difference between $\rho^{\text{QMC}}(t)$ and $\rho^{\text{exact}}(t)$, given that $\rho^{\text{exact}}(t)$ is positive semidefinite. This approach is valid in the sense that $\rho^{\text{QMC}}(t)$ is numerically close to $\rho^{\text{exact}}(t)$ element-wise. Note that this differs from the $L^2$ norm discussed in the previous section, which is not element-wise. In Fig. S3(b), we plot $\|\rho^{\text{QMC}} - \rho^{\text{exact}}\|_F$ as a function of circuit time for the positivity check. Similar to the Hermitian check, using a larger walker number leads the QMC-approximated density matrix deviating less from a positive semidefinite matrix, and this deviation grows only mildly over time.

Based on the numerical results in this section, the QMC method offers the convenient feature of needing only one control parameter, $N^{\text{diag}}$. This parameter balances simulation accuracy and robustness, which is improved by increasing the number of walkers, against computational cost, which is reduced by using fewer walkers. Moreover, this balance can be monitored simply by checking whether $N^{\text{diag}}(t)$ deviates from its initial value $N^{\text{diag}}$, a cost-free procedure, while other convergence properties automatically follow once $N^{\text{diag}}(t)$ remains close to $N^{\text{diag}}$.

## S4. HILBERT SPACE TRUNCATION AND PSEUDO-SPARSITY OF THE DENSITY MATRIX

In the main text, we briefly mentioned that our QMC algorithm adopts the FCIQMC initiator approximation [4, 6], which we refer to here as *stochastic truncation*. Together with *stochastic compression*, this approximation enables QMC to achieve both memory and speed advantages over exact master-equation and QT solvers by working in an $O(D)$-dimensional space while still providing an accurate approximation of the exact density matrix after averaging. Specifically, we define the QMC Hilbert space as the set of all states occupied by QMC walkers,

$$\{\text{loc} \mid \text{loc} = \text{loc}_\alpha \text{ for some } \alpha \in \{1, \ldots, N_t^{\text{tot}}\}\}, \tag{S24}$$

and its dimension by

$$\dim(\mathcal{H}_{\text{QMC}}) := \left|\{\text{loc} \mid \text{loc} = \text{loc}_\alpha \text{ for some } \alpha \in \{1, \ldots, N_t^{\text{tot}}\}\}\right|, \tag{S25}$$

which, when stochastic compression and truncation are effective, is $O(D)$——much smaller than the full $D^2$-dimensional space. As stated in the main text, this is because (a) the noise effectively reduces the magnitude of the off-diagonal elements over time, which allows compression and truncation and (b) even for weak noise, moderate over-truncation only introduces minimal bias once multiple samples are averaged. In this section, we present numerical evidence supporting (b).

We first detail how to perform this truncation numerically. Since the innermost loop of Algorithm 1 is the most computationally intensive part of our QMC algorithm, reducing $\dim(\mathcal{H}_{\text{QMC}})$ improves its computational scaling. This approach is justified because states with small populations not only contribute negligibly to measurements but also have minimal impact on the incremental density vector $|\Delta\rho(t)\rangle\rangle$ after the time propagation, $\Delta t \mathcal{L}|\rho(t)\rangle\rangle$. Consequently, we perform a state truncation such that a state $|i,j\rangle\rangle$ whose walker number is below a certain threshold, $N_{ij}(t) \leq \xi N^{\text{diag}}$, is deemed a non-initiator and cannot expand the QMC Hilbert space by spawning walkers into unoccupied states. In other words, if a newly spawned walker $w_{t+\Delta t}^{(\alpha')}$ originates from a non-initiator state and $\text{loc}_{\alpha'}$ was unoccupied at time $t$, then $w_{t+\Delta t}^{(\alpha')}$ is discarded. Algorithm 2 summarizes this updated scheme. Note that as the circuit evolves, non-initiator states may become initiators, and vice versa.

Next, we examine how different truncation ratios $\xi$ affect both simulation efficiency and accuracy. In Fig. S4, we perform 10-qubit and 12-qubit GHZ state preparation simulations (as in the main text) while varying $\xi$, and plot $\dim(\mathcal{H}_{\text{QMC}})$ against $\xi$. We observe that $\dim(\mathcal{H}_{\text{QMC}})$ decreases sharply for $\xi > 0.01\%$, then eventually saturates at a level where initiator states no longer spawn walkers into unoccupied states, preventing further growth of the QMC Hilbert space. In the inset, we use the 10-qubit case, where an exact solution is available, to compute the trace-norm distance

$$T(\rho, \sigma) = \frac{1}{2}\|\rho - \sigma\|_1$$

between the QMC-approximated and exact density matrices. For each $\xi$, we run five individual samples and determine both the average trace distance over those runs (green circles) and the trace distance of the averaged density matrix (blue circles). The green circles illustrate how large $\xi$ (over-truncation) degrades accuracy on individual runs, whereas the blue circles show that averaging helps mitigate this loss. Empirically, $\xi = 0.1\%$ (the value used in all simulations) is a sweet spot, where $\dim(\mathcal{H}_{\text{QMC}})$ is reduced to $O(D)$, thereby significantly enhancing efficiency, while retaining trace-norm distances more comparable to smaller $\xi$ values than to larger ones, thus preserving accuracy.

We then investigate whether $\dim(\mathcal{H}_{\text{QMC}}) = O(D)$ continues to hold for larger system sizes across different circuits under $\xi = 0.1\%$. In Fig. S5, we plot $\dim(\mathcal{H}_{\text{QMC}})$ versus the number of qubits for all circuits examined in the main text. For small systems, the QMC Hilbert space remains nearly as large as the full $D$, offering little advantage over an exact solver. However, as the system grows, stochastic compression and truncation cause $\dim(\mathcal{H}_{\text{QMC}})$ to grow far more slowly than $D^2$. A linear fit on a log scale of the last four data points reveals exponents of $2.12^n$ and $1.81^n$ for the crosstalk-suppression circuits, and $1.93^n$ for the GHZ state preparation circuits, numerically verifying the $O(D)$ scaling. Although the QMC Hilbert space is still slightly larger than that of a single-trajectory QT solver in these benchmarks, QT would require tens of thousands of trajectories to match QMC's error bars (QMC needs fewer than ten samples). This substantial difference in convergence speed underlies the computational advantage of QMC over QT.

Lastly, we show that the Liouvillian matrix $\mathcal{L}(t)$ in our model is sparse, and that its sparsity grows exponentially with system size. Recall that

$$\mathcal{L}(t) = -i\mathbb{I} \otimes H(t) + iH^{\text{T}}(t) \otimes \mathbb{I} + \frac{1}{2}\sum_k \gamma_k(t)(2L_k \otimes L_k - \mathbb{I} \otimes L_k^\dagger L_k - L_k^\dagger L_k \otimes \mathbb{I}).$$

The non-zero elements in $\mathcal{L}(t)$ can be bounded using the triangle inequality as

$$\|\mathcal{L}(t)\|_0 \leq 2\|\mathbb{I} \otimes H(t)\|_0 + \sum_k \left(\|L_k \otimes L_k\|_0 + \|\mathbb{I} \otimes L_k^\dagger L_k\|_0 + \|L_k^\dagger L_k \otimes \mathbb{I}\|_0\right)$$
$$= 2D\|H(t)\|_0 + \sum_k \left(\|L_k\|_0^2 + 2D\|L_k^\dagger L_k\|_0\right). \tag{S26}$$

Here, we consider the extreme case where each circuit layer has a maximal number of two-qubit gates. Each entangling gate contributes $D$ nonzero elements to the Hamiltonian matrix; combining this with the intrinsic qubit frequency and crosstalk terms, the maximum number of non-zero elements in $H(t)$ is $(1 + n/2)D$, i.e., $\|H(t)\|_0 \leq (1 + n/2)D$.

---

**Algorithm 2:** Real-time Open QMC Algorithm with Truncation

---

**Initialization:** Set the diagonal walker number $N^{\text{diag}}$, the truncation ratio $\xi$, and construct the initial population vector based on the given initial state:
$$|N(0)\rangle\!\rangle := N^{\text{diag}}|\rho(0)\rangle\!\rangle.$$

**for** $n = 1 : n_{\text{sample}}$ **do**
    **for** $t = 0 : \Delta t : t_f$ **do**
        Initialize an empty set for spawned walkers: $\{w^{(\alpha)}_{t+\Delta t}\}_\alpha$.
        **for** *each occupied state* $\text{loc}_k = |i,j\rangle\!\rangle$ *with* $N_{ij}(t)$ *walkers* **do**
            **Spawning:**
            Identify all states $|k,l\rangle\!\rangle$ connected to $|i,j\rangle\!\rangle$ and compute $p^{\text{sp}}_{ij}(t)$ as in Eq. (S9).
            Use binomial sampler with sample size $N_{ij}(t)$ and success rate $p^{\text{sp}}_{ij}(t)$:
$$N^{\text{sp}}_{ij}(t) \leftarrow B(N_{ij}(t), p^{\text{sp}}_{ij}(t))$$
            Fill the spawned walker set with the multinomial distribution defined in Eq. (S10):
$$\{w^{(\alpha)}_{t+\Delta t}\}_\alpha \leftarrow \text{Multinomial}(N^{\text{sp}}_{ij}(t), p_{ij\to(kl,c)})$$
            Assign proper signs to the spawned walkers
            **for** *each spawned walker* $w^{(\alpha')}_{t+\Delta t}$ **do**
                **if** $N_{ij}(t) \leq \xi N^{\text{diag}}$ *and* $\text{loc}_{\alpha'}$ *is unoccupied* **then**
                    discard this walker
                **else**
                    keep this walker in the spawn set
                **end**
            **end**
        **end**
        **Annihilation:** Merge the spawned walkers into the population vector:
$$|N(t+\Delta t)\rangle\!\rangle \leftarrow |N(t)\rangle\!\rangle + \sum_\alpha w^{(\alpha)}_{t+\Delta t},$$
        where pairs of walkers at the same location with opposite signs ($\pm 1$ or $\pm i$) are annihilated and removed. The diagonal walker population is updated using the spawned walkers occupying the diagonal elements of the density matrix:
$$N^{\text{diag}}_{t+\Delta t} \leftarrow N^{\text{diag}}_t + \left\|\text{Re}\left(\text{diag}(\{w^{(\alpha')}_{t+\Delta t}\})\right)\right\|_0 + i \cdot \left\|\text{Im}\left(\text{diag}(\{w^{(\alpha')}_{t+\Delta t}\})\right)\right\|_0.$$
        **Measurement:** If a measurement is required at $t + \Delta t$, compute the expectation value with the estimator:
$$\text{tr}\left(O \frac{1}{N^{\text{diag}}_{t+\Delta t}}|N(t+\Delta t)\rangle\!\rangle\right).$$
    **end**
**end**

---

For the amplitude damping channel, $L_k = I^{\otimes(k-1)} \otimes \sigma_- \otimes I^{\otimes(n-k)}$, we have

$$L_k^\dagger L_k = I^{\otimes(k-1)} \otimes (\sigma_+ \sigma_-) \otimes I^{\otimes(n-k)} = I^{\otimes(k-1)} \otimes \begin{pmatrix} 1 & 0 \\ 0 & 0 \end{pmatrix} \otimes I^{\otimes(n-k)},$$

and

$$\|L_k\|_0 = \frac{D}{2}, \|L_k^\dagger L_k\|_0 = \frac{D}{2}.$$

Similarly, for the dephasing channel, $L_k = I^{\otimes(k-1)} \otimes \sigma_z \otimes I^{\otimes(n-k)}$, we have

$$\|L_k\|_0 = D, \|L_k^\dagger L_k\|_0 = \|I^{\otimes(k-1)} \otimes \begin{pmatrix} 1 & 0 \\ 0 & 1 \end{pmatrix} \otimes I^{\otimes(n-k)}\|_0 = D.$$

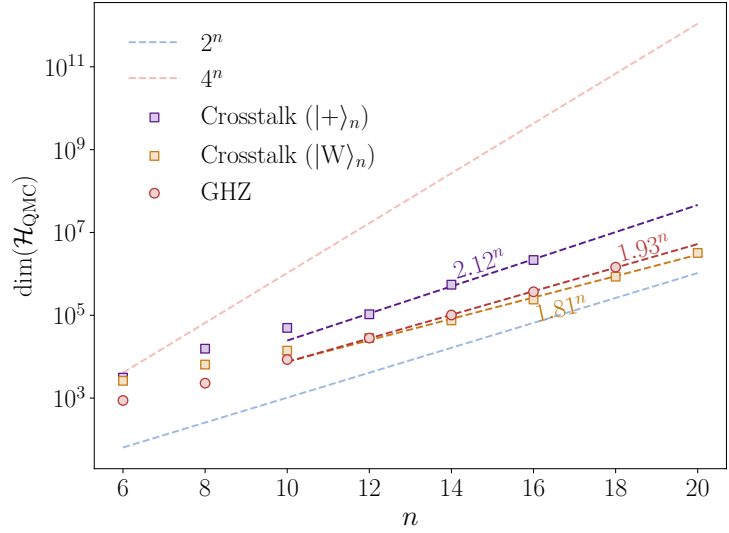

FIG. S5. $\dim(\mathcal{H}_{\text{QMC}})$ plotted against the number of qubits for all circuits benchmarked in the main text. The final four data points in each case are fitted linearly on a log scale. Dashed lines show the scaling for the exact master equation solver ($D^2 = 4^n$) and a single-shot QT solver ($D = 2^n$). All simulations use a truncation ratio of $\xi = 0.1\%$.

Combining all of the above leads to

$$\|\mathcal{L}(t)\|_0 \leq (2+n)D^2 + n(D^2 + 2D^2 + \frac{D^2}{4} + D^2) = O(nD^2) \tag{S27}$$

Since the dimension of $\mathcal{L}(t)$ is $D^2 \times D^2$, the sparsity of $\mathcal{L}(t)$ is

$$\frac{\|\mathcal{L}(t)\|_0}{D^4} = O(\frac{1}{D^2}) = O(\frac{1}{4^n}), \tag{S28}$$

i.e., becomes exponentially sparse as the system size increases. Moreover, each column has at most $O(n)$ nonzero elements. Since QMC only requires storing the columns corresponding to walker-occupied states and operates in a space of dimension $O(D)$, at most $O(nD)$ elements need to be stored in memory—far fewer than the $O(D^4)$ total elements in $\mathcal{L}(t)$.

## S5. CIRCUIT SIMULATION DETAILS

In this section, we provide a detailed account of how the quantum circuits in the main text are implemented numerically. We first explain how the basis transformation is performed when the initial state is the maximally coherent $|+\rangle_n$ state for the crosstalk suppression simulation. Since in this case the density matrix in the computational basis is full and the initial population vector must be initialized with all $D^2$ states occupied, it is not feasible to store the latter even for moderate sizes. This is so despite many of the states eventually decaying to zero due to decoherence. To circumvent this, we rotate into the Pauli-$X$ basis via a collective Hadamard transform:

$$|+\rangle_n \mapsto |0\rangle_n, \sigma_x \mapsto \sigma_z, \sigma_z \mapsto \sigma_x, \sigma_- \mapsto \sigma_x + i\sigma_y. \tag{S29}$$

Under this transformation, the initial $|+\rangle_n$ state becomes the sparse $|0\rangle_n$ state, crosstalk terms transform into $XX$ couplings, and the dynamical decoupling sequence transforms into a staggered $ZZ$ sequence. Meanwhile, relaxation noise takes the form $\sigma_x + i\sigma_y$, which reintroduces coherence over time and reduces sparsity in the density matrix. Consequently, the $|+\rangle_n$ case is the most challenging scenario for QMC, as reflected by the increasing total walker number reported in the main text. Nevertheless, by working in this rotated basis, we only need to initialize our QMC simulation in a significantly smaller effective space $\dim(\mathcal{H}_{\text{QMC}})$, as illustrated in Fig. S5.

We next illustrate the pulse-level realization of gates. As shown in Fig. S6, we use a 4-qubit system as an example and larger systems can be generalized automatically. All gates are realized as square pulses of constant frequency.

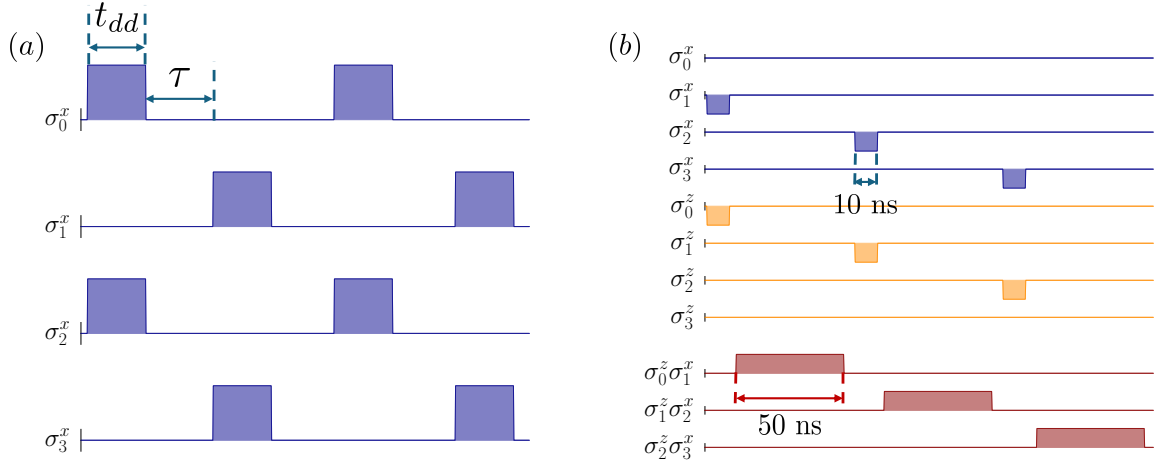

FIG. S6. Control pulse of the circuits simulated in the main text for (a) crosstalk suppression and (b) GHZ state preparation. Each line represents one or more control pulses acting on a specific qubit. The pulse width represents the duration. Time flows from left to right.

The pulse of a single-qubit $X$ gate in Fig. S6(a) that rotates qubits through $\pi$ radians around the x-axis has the Hamiltonian $H_{x,i} = \omega_{dd}\sigma_i^x$, where $\omega_{dd}$ is the angular pulse frequency (in units of $2\pi$), and the evolution operator over the duration of $t_{dd}$ is $\exp(-it_{dd}H_{x,i}) = \exp\left(-i\frac{\pi}{2}\sigma_i^x\right) = -i\sigma_i^x$ with an unobservable global phase $-i$, and we set $\omega_{dd}t_{dd} = \frac{1}{4}$. In our simulations, $t_{dd} = 10$ ns for the $|+\rangle_n$ state and $t_{dd} = 2$ ns for the $|W\rangle_n$, because the latter is an entangled state and requires a higher pulse frequency to suppress the crosstalk. The idle time $\tau$ is determined by the number of DD sequences and the total circuit time as

$$\tau = \frac{\text{total circuit time}}{\text{\# of DD sequences} \times 4} - t_{dd}, \quad (S30)$$

where the factor of 4 accounts for the four operations in a single staggered $XX$ cycle (two $X$-pulses of duration $t_{dd}$ and two idle periods $\tau$). The gates in Fig. S6(b) for GHZ state preparation circuit are implemented similarly, with gate Hamiltonians $H_{x,i} = \omega_x \sigma_i^x$, $H_{z,i} = \omega_z \sigma_i^z$ and $H_{zx,i} = \omega_{zx}\sigma_i^z\sigma_{i+1}^x$ for $X$-rotation, $Z$-rotation and $ZX$-coupling, respectively. There is no idle time in this circuit and since the duration of two-qubit gates (50 ns) is longer than that of the single-qubit gate (10 ns), we set $\omega_x = \omega_z = 5\omega_{zx}$.

Lastly, for GHZ state preparation, we realize a CNOT gate through the following three-step sequence: (1) $R_z(-\frac{\pi}{2})$ on the control qubit, (2) $R_x(-\frac{\pi}{2})$ on the target qubit and (3) $R_{zx}(\frac{\pi}{2})$. We show that this composite operation is equivalent to a CNOT up to a global phase. Denoting the three matrices as:

$$\text{The control pre-rotation: } C = R_z\left(-\frac{\pi}{2}\right) \otimes I = \begin{pmatrix} e^{i\pi/4} & 0 \\ 0 & e^{-i\pi/4} \end{pmatrix} \otimes I = \begin{pmatrix} e^{i\pi/4} & 0 & 0 & 0 \\ 0 & e^{i\pi/4} & 0 & 0 \\ 0 & 0 & e^{-i\pi/4} & 0 \\ 0 & 0 & 0 & e^{-i\pi/4} \end{pmatrix}$$

$$\text{The target pre-rotation: } T = I \otimes R_x\left(-\frac{\pi}{2}\right) = I \otimes \frac{1}{\sqrt{2}}\begin{pmatrix} 1 & i \\ i & 1 \end{pmatrix} = \frac{1}{\sqrt{2}}\begin{pmatrix} 1 & i & 0 & 0 \\ i & 1 & 0 & 0 \\ 0 & 0 & 1 & i \\ 0 & 0 & i & 1 \end{pmatrix}$$

$$\text{The two-qubit entangling gate: } E = \exp\left(-i\frac{\pi}{4}(Z \otimes X)\right) = \frac{1}{\sqrt{2}}\begin{pmatrix} 1 & -i & 0 & 0 \\ -i & 1 & 0 & 0 \\ 0 & 0 & 1 & i \\ 0 & 0 & i & 1 \end{pmatrix}.$$

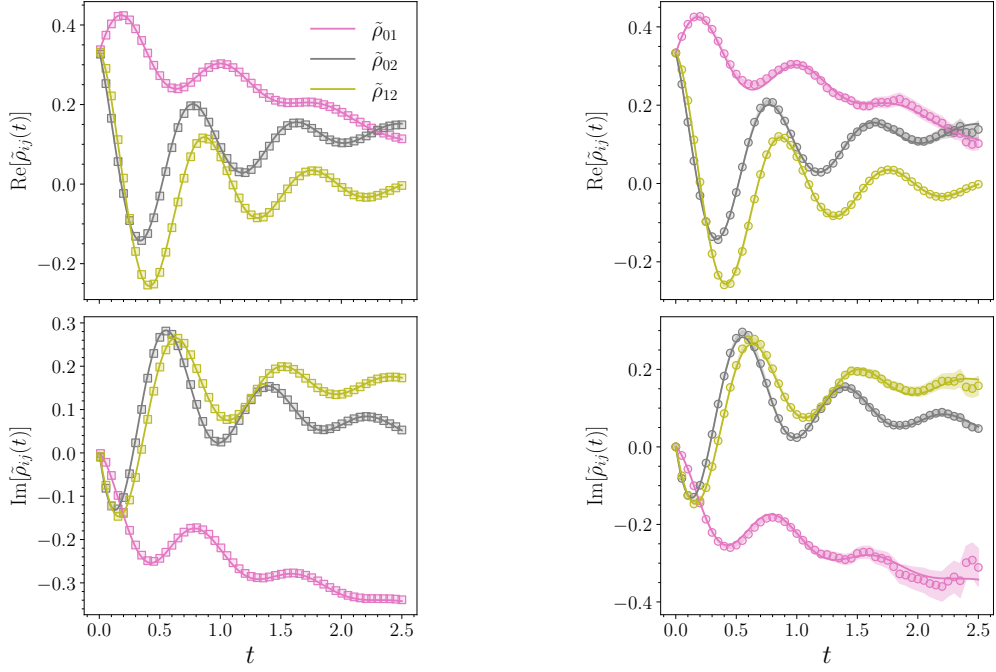

FIG. S7. Non-Markovian dynamics for the real and imaginary part of off-diagonal density matrix elements $\tilde{\rho}_{01}$, $\tilde{\rho}_{02}$ and $\tilde{\rho}_{12}$. Solid lines represent exact master equation solutions in both panels. Left: Results from a single QMC sample with $10^6$ walkers. Right: QT solutions averaged over $10^4$ trajectories.

The overall operator is

$$U = CTE = \begin{pmatrix} e^{i\pi/4} & 0 & 0 & 0 \\ 0 & e^{i\pi/4} & 0 & 0 \\ 0 & 0 & 0 & i\,e^{-i\pi/4} \\ 0 & 0 & i\,e^{-i\pi/4} & 0 \end{pmatrix}.$$

Including a global $e^{-i\pi/4}$ phase then yields the standard CNOT.

## S6. MORE ON NON-MARKOVIAN DYNAMICS

In addition to tracking the diagonal elements, we extended our simulations to include the off-diagonal dynamics in the same two-qubit model used in the main text. As before, we work in the rotated orthonormal basis defined by the Lindblad operators and employ the Redfield master equation with negative jump rates. The results (see Fig. S7) reveal that, for QT, the real parts of the off-diagonal elements diverge only weakly at later times, while the imaginary parts exhibit a more pronounced divergence. In contrast, QMC yields off-diagonal elements that remain much closer to the exact solution, with both the real and imaginary components converging effectively. These findings further underscore QMC's convergence advantage in capturing the non-Markovian dynamical features of open quantum systems.


\end{document}